\documentclass[superscriptaddress,twocolumn,showpacs]{revtex4-1}
\usepackage[utf8]{inputenc} 
\usepackage{amsmath,amsthm}
\usepackage{mathtools}
\usepackage{braket}
\usepackage{graphicx}
\usepackage{amsfonts}
\usepackage{pgfplots}
\usepackage{csquotes}
\usepackage{hhline}
\usepackage{amssymb}
\usepackage{listings}
\usepackage{color}
\usepackage{comment}
\definecolor{codegreen}{rgb}{0,0.6,0}
\definecolor{codegray}{rgb}{0.5,0.5,0.5}
\definecolor{codepurple}{rgb}{0.58,0,0.82}
\definecolor{backcolour}{rgb}{0.95,0.95,0.92}
 
\lstdefinestyle{mystyle}{
    backgroundcolor=\color{backcolour},   
    commentstyle=\color{codegreen},
    keywordstyle=\color{magenta},
    numberstyle=\tiny\color{codegray},
    stringstyle=\color{codepurple},
    basicstyle=\footnotesize,
    breakatwhitespace=false,         
    breaklines=true,                 
    captionpos=b,                    
    keepspaces=true,                 
    numbers=left,                    
    numbersep=5pt,                  
    showspaces=false,                
    showstringspaces=false,
    showtabs=false,                  
    tabsize=2
}
\theoremstyle{plain}
\newtheorem{thm}{Theorem}[section]
\usepackage{subfigure}

\usepackage{dcolumn}
\usepackage{tabularx}
\usepackage{float}

\usepackage[colorlinks=true,linkcolor=blue,citecolor=blue,urlcolor=blue]{hyperref}
\usepackage{longtable}
\usepackage{braket}
\usepackage{float}
\newcolumntype{C}{>{\centering\arraybackslash}X}
\begin{document}
\title{Demonstration of a general fault-tolerant quantum error detection code for $(2n+1)$-qubit entangled state on IBM 16-qubit quantum computer}

\author{Ranveer Kumar Singh}
\email{ranveersfl@gmail.com}
\affiliation{Department of Mathematics, \\Indian Institute of Science Education and Research Bhopal, Bhauri 462066, Madhya Pradesh, India}
\author{Bishvanwesha Panda}
\email{bishvanweshapanda@gmail.com}
\affiliation{Indian Institute of Science Education and Research Kolkata,\\ Mohanpur 741246, West Bengal, India}

\author{Bikash K. Behera}
\email{bikash@bikashsquantum.com}
\affiliation{Bikash's Quantum (OPC) Private Limited, Balindi, Mohanpur, 741246, Nadia, West Bengal, India}
\affiliation{Department of Physical Sciences,\\ Indian Institute of Science Education and Research Kolkata, Mohanpur 741246, West Bengal, India}
\author{Prasanta K. Panigrahi}
\email{pprasanta@iiserkol.ac.in}
\affiliation{Department of Physical Sciences,\\ Indian Institute of Science Education and Research Kolkata, Mohanpur 741246, West Bengal, India}

\begin{abstract}
Quantum error detection has always been a fundamental challenge in a fault-tolerant quantum computer. Hence, it is of immense importance to detect and deal with arbitrary errors to efficiently perform quantum computation. Several error detection codes have been proposed and realized for lower number of qubit systems. Here we present an error detection code for a $(2n+1)$-qubit entangled state using two syndrome qubits and simulate it on IBM's 16-qubit quantum computer for a 13-qubit entangled system. The code is able to detect an arbitrary quantum error in any one of the first $2n$ qubits of the $(2n+1)$-qubit entangled state and detects any bit-flip error on the last qubit of the $(2n+1)$-qubit entangled state via measurements on a pair of ancillary error syndrome qubits. The protocol presented here paves the way for designing error detection codes for the general higher number of entangled qubit systems.    
\end{abstract}

\begin{keywords}{IBM Quantum Experience, Quantum Error Detection, Entangled States}\end{keywords}

\maketitle

\section{Introduction}
Quantum errors are the inevitable obstacles for realizing a fault-tolerant quantum computer \cite{qed_PreskillPhysicstoday1999,qed_ShorIEEE1996,qed_SteaneNat1999}. Quantum systems show much more pronounced noise effects on them through quantum errors. While classical computers are only affected by bit-flip errors, quantum computers exhibit mainly three types of errors such as bit-flip, phase-flip and arbitrary phase-change error \cite{qed_NiggScience2016,qed_GhoshQIP2018}. Thus fault tolerant quantum computation projects a daunting task to accomplish. In order to run quantum algorithms with large time complexity, improvement needs to be done using quantum error correction protocols \cite{qed_ChiaveriniNature2004,qed_GottesmanarXiv:97050521997} and fault tolerant schemes \cite{qed_RaussendorfPRL2007,qed_ChildressPRL2006}. Several experiments have already been performed to demonstrate the usefulness of quantum error correcting codes to protect a quantum memory \cite{qed_KnillPRL2001}. To implement an error correction code, detection of error is needed, hence becoming an important part of error correction scheme. Several error detection  as well as correction codes have been proposed \cite{qed_Niggscience2014,qed_LinkeScienceAdvances2017,qed_TakitaPRL2017,qed_FarkasIEEE1995,qed_FeldmeierIEEE1995,qed_McAuleyIEEE1994,qed_NguyenCISS1998,qed_GuptaIJQI2005}. The pioneering work on error detection as well as correction had been started by Shor \cite{qed_ShorPRA1995} and Steane \cite{qed_SteanePRL1996,qed_SteanePRSL1996,qed_SteaneIEEE1999}. Since then quantum error detection and error correction have been a subject of intense study. 

Recently, Corcoles \textit{et al.} \cite{qed_CorcolesNatComm2015} proposed a quantum error detection code for one of the Bell states using two ancillary syndrome qubits and demonstrated it experimentally using a square lattice structure of four superconducting qubits. In the proposed error detection code, they used a two-by-two lattice structure i.e., the square lattice of superconducting qubits. They verify the non-demolition nature of the protocol by demonstrating the preservation of entangled state through high fidelity syndrome measurements in the presence of an arbitrary applied error. The surface code (SC) \cite{qed_BravyiarXiv9811052,qed_KitaevAP1997} has emerged as a promising candidate for quantum computers based on superconducting qubits due to its nearest-neighbour qubit interaction and high fault-tolerant error thresholds \cite{qed_RaussendorfPRL2007}. In recent times, superconducting qubits have become potential candidates for the realization of SC \cite{qed_FowlerPRA2012,qed_ChowNatcommun2014} with continuous improvement in coherence times \cite{qed_PaikPRL2011,qed_ChuangAPL2013,qed_BarendaPRL2013} and quantum errors \cite{qed_BarendaNat2014}. A highly efficient new quantum computer has been developed by IBM which uses superconducting transmon based qubits for computing. IBM quantum computer has become a completely new candidate for the implementation of SC. Qubits of the IBM quantum computers are placed at the vertices of a two dimensional array. A large number of works have been performed by researchers using IBM quantum computers. 

Recently Debjit \textit{et al.}\unskip~\cite{qed_GhoshQIP2018} experimentally realized an error correction code for Bell state and GHZ state on IBM 5-qubit quantum computer and generalized it to n-qubit case. IBM quantum experience from its inception has gained a lot of popularity in the research community since the cloud based access provided by IBM has been used to accomplish various tasks \cite{qed_AggarwalarXiv:1804.08655v12018,qed_SrinivasanarXiv:1805.109282018,qed_BeheraQIP2017,qed_GangopadhyayQIP2017,qed_HegadearXiv:1712.073262017,qed_BeheraarXiv:1712.008542017,qed_HarperarXiv:1806.023592018,qed_KlcoarXiv:1803.033262018,qed_DasharXiv1710.051962017,qed_VishnuarXiv:1709.05697,qed_SatyajitarXiv:1712.05485,qed_RoyarXiv:1710.10717,qed_KalraarXiv:1707.09462,qed_JhaarXiv:1806.10221,qed_DasharXiv:1805.10478,qed_BeherarXiv:1803.06530,qed_SrinivasanarXiv:1801.00778,qed_GurnaniarXiv:1712.10231,qed_KapilarXiv:1807.00521,qed_MohantaarXiv:1807.00323,qed_BeherarXiv:1806.10229}. Thus testing and implementing error detection codes using IBM quantum computers opens up new horizons of research. It has been shown that error detection is very useful on IBM 5Q chips \cite{qed_VuillotarXiv:1705.089572017}. Error detection and correction remains a challenging problem for arbitrary entangled states with large number of qubits. Although several error detection codes have made good amount of progress in the pursuit, still much progress needs to be made.

We take the study a step forward and propose an error detecting code for a $(2n+1)$-qubit entangled state prepared from a $2n$-qubit entangled state possessing a kind of ``complementarity" property which will be explained in detail in Section \ref{qed_section2}. Simply, an entangled state has the  complementarity property if for every term appearing in the state there is another term in the state complementary to it, where the complementary state is obtained by a modular sum with 1 to each of the qubit. For example $\Ket{11101}$ is the complementary state to $\Ket{00010}$. The entangled states with this complementarity property are general and cover the maximally entangled state Bell states and all the generalized GHZ states. The proposed protocol is useful as it can be used to detect errors in generalized GHZ states with even number of qubits and all Bell states which are used in many quantum algorithms as quantum teleportation \cite{qed_BennettPRL1993}, quantum cryptography \cite{qed_BennettIEEE1984}, quantum key distribution \cite{qed_BennettJC1992}, quantum secret sharing \cite{qed_HilleryarXiv98060631998}, superdense coding \cite{qed_BennettPRL1992} etc. 

In our protocol, we first take any $2n$-qubit entangled state with complementarity property and add another qubit to the state using CNOT operations as the result of which we get a $(2n+1)$-qubit entangled state depending on the terms in the state of the $2n$-qubit state taken. We then add two error syndrome qubits to the state prepared above in a way such that they remain in a product state. Then measurement is performed on the syndrome qubits and depending on the result of the measurement, we conclude the type of error present in the $(2n+1)$-qubit entangled state. Our protocol detects any arbitrary single-qubit phase-change error or bit-flip or phase-flip error on any of the $2n$ qubits and detects only bit-flip error in the last qubit of the $(2n+1)$-qubit state. To demonstrate the usefulness of the protocol, we perform a simulation with a 13-qubit entangled state on the IBM 16-qubit quantum computer and compare the results for various types of errors. We implement the errors on the qubits using different gates provided by IBM quantum experience. We design the quantum circuit using QASM language and simulate it using QISKit. 

\section{Results}
\label{qed_section2}

\begin{figure*}[]
\centering
\includegraphics[width=\linewidth]{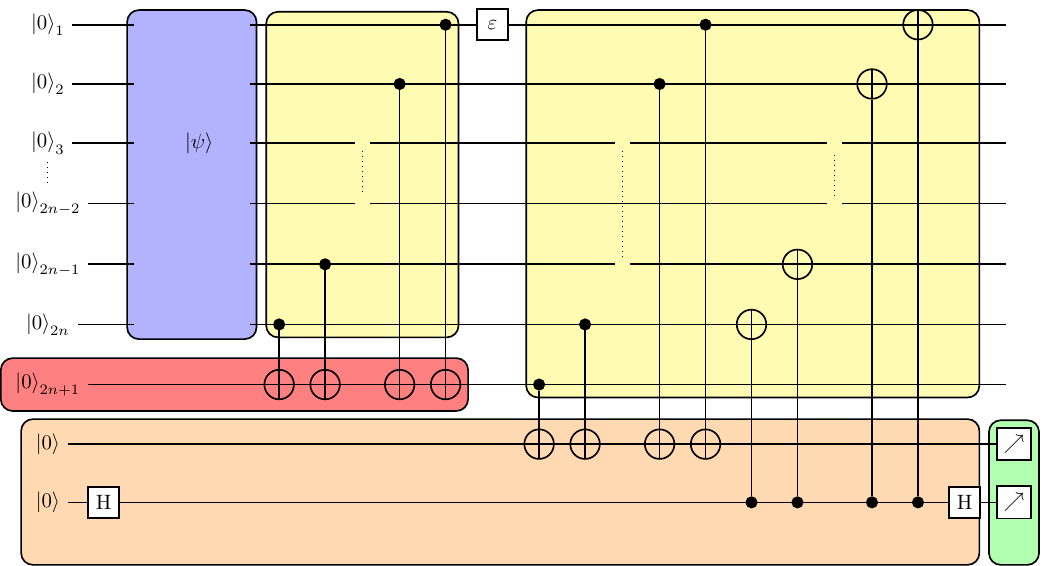}
\caption{\normalsize{\textbf{Quantum circuit for quantum error detection in a $(2n+1)$-qubit entangled state.} The blue box generates $2n$-qubit entangled state. The red box contains the additional qubit to be added to the $2n$-qubit entangled state. The smaller yellow box depicts the circuit used to entangle an additional qubit to the $2n$-qubit entangled state to prepare a $(2n+1)$-qubit entangled state. The orange box contains two syndrome qubits to be used to detect quantum errors in the $(2n+1)$-qubit entangled state. The larger yellow box depicts the circuit used to add the syndrome qubits to the $(2n+1)$-qubit entangled state for error detection. The green box contains two measurement operations to be made on the two syndrome qubits. The error $\varepsilon$ is introduced in qubit 1 and detected from the results of the measurement on the two syndrome qubits. The circuit works equally well for any arbitrary error introduced in any single qubit from 1 to $2n$ and bit-flip error in qubit $2n+1$.}}
\label{qed_fig1}
\end{figure*}

Our circuit consists of a entangled state of $(2n+1)$ number of qubits and two syndrome qubits. We first prepare a $2n$ qubit entangled state of a special form as outlined here. Let $A_n$ be the set of computational basis states for $2n$ qubits, that is, $A_n=\{\Ket{a_1a_2\dots a_{2n}} ; ~a_i=0,1~ \forall~ i=1,2,\dots 2n\ , n\in \mathbb{N}\}$ of state vectors, where the first number in the ket represents the first qubit, the second number represents the second qubit and so on. Let $B_n$ be a nonempty subset of $A_n$ with the property that if $\Ket{a_1a_2\dots a_{2n}}\in B_n$ then $\Ket{(a_1\oplus1)(a_2\oplus1)\dots(a_{2n}\oplus1)}\in B_n$, where $\oplus$ is addition modulo 2. For brevity, we call the ket\\ $\Ket{(a_1\oplus1)(a_2\oplus1)\dots(a_{2n}\oplus1)}$ complementary to $\Ket{a_1a_2\dots a_{2n}}\in B_n$ and the set $B_n$ a set with ``complimentarity property". Now consider the sum \begin{equation}
\begin{split}
    \Ket{\psi}_{B_n}= \frac{1}{\sqrt{|B_n|}}\sum_{\Ket{a_1a_2\dots a_{2n}}\in B_n}{}\pm\Ket{a_1a_2\dots a_{2n}}
\label{qed_Eq.1}
\end{split}
\end{equation}

where $|B_n|$ is the cardinality of the set $B_n$. We call such states as states with ``complementarity property". We will prove later that $\Ket{\psi}_{B_n}$ is entangled when $B_n$ is a proper subset of $A_n$ (Theorem \ref{thm III.1}). For example one state of the above form with 4 qubits could be $\Ket{\psi}=\frac{1}{\sqrt{6}}\big(\Ket{0000}+\Ket{1111}+\Ket{1010}+\Ket{0101}+\Ket{0111}+\Ket{1000}\big)$. Note that $A_n$ trivially has the complementarity property. So in case $A_n=B_n$, there are special states called ``graph states" which are also entangled (see Section \ref{qed_section3} for details). We then add another qubit to this $2n$ qubit state by CNOTs as shown in Fig. \ref{qed_fig1}. The resultant state is still entangled (see Theorem \ref{thm III.2} for proof) depending on the terms in the sum. If the $(2n+1)$th qubit does not get entangled to starting $2n$-qubit entangled state then the detection of error in the $2n$-qubit entangled state is trivially done by the circuit. Thus, the protocol detects any quantum error in Bell states and generalized GHZ states as one of the special cases. After the preparation of $(2n+1)$-qubit entangled state, two syndrome qubits are added in the circuit as shown in the Fig. \ref{qed_fig1}. Then measurement is done on the error syndrome qubits to detect errors. The above Table \ref{qed_table1} summarizes the results of measurement and the type of error (for a detailed discussion on measurement results, Methods section can be checked).

\begin{table}[]
\begin{center}
 \begin{tabular}{c c} 
 \hline
 \hline
 Measurement result & Type of error  \\ [0.5ex] 
 \hline
 \hline
 $\Ket{00}$ & No error  \\ 
 \hline
 $\Ket{10}$ & Bit-flip in any one qubit.  \\
 \hline
 $\Ket{01}$ & Phase-flip in any one qubit. \\
 \hline
 $\Ket{11}$ & Bit-flip and phase-flip in any one qubit.     \\ [1ex] 
 \hline
 \hline
\end{tabular}
\caption{\textbf{Measurement results of syndrome qubits and type of errors}}
\label{qed_table1}
\end{center}
\end{table}

The circuit presented here detects any quantum error present in any single qubit from qubit 1 to qubit $2n$ and detects any bit-flip error in $(2n+1)$th qubit. 

\subsection{Implementation of the error detection protocol.}
We demonstrate the quantum error detection protocol by simulating the circuit in Fig. \ref{qed_fig1} for a 13-qubit entangled state prepared using a $12$-qubit graph state and adding another qubit using CNOT operations  on the \textit{ibmqx5} quantum computer (see Section \ref{qed_section3} for details). We apply single-qubit rotations to first qubit in the 13-qubit entangled state with the form $\varepsilon=R_{\theta}$ where $R$ defines the rotation axis and $\theta$ is the angle of rotation. We choose to apply the error on the first qubit but errors can also be introduced in any of the 12 qubits with the exception of 13th qubit where only bit-flip error can be detected. In the first case, we introduce no error i.e., we apply $\varepsilon=R_{0}$ on the first qubit. In this case, both the syndrome qubits are measured to be in their ground state $\Ket{0}$. Next, for bit-flip error we apply $\varepsilon=X_{\pi}$ on the first qubit. In this case, the first error syndrome qubit gets excited to $\Ket{1}$. In case of phase-flip error, $\varepsilon=Z_{\pi}$ is applied on the first qubit and the first and the second ancillary syndrome qubit is measured to be in $\Ket{0}$ and $\Ket{1}$ state respectively. The measurement results remain the same irrespective of the 12 qubits on which the error is applied. In case of a bit-flip error on the 13th qubit, the measurement result remains same i.e., $\Ket{10}$ where it is understood that the first number in the ket represents the first qubit and the second number represents the second qubit. Thus the simulation result confirms Table \ref{qed_table1}.\\\\
\begin{figure*}
\centering
\includegraphics[width=0.65\textwidth]{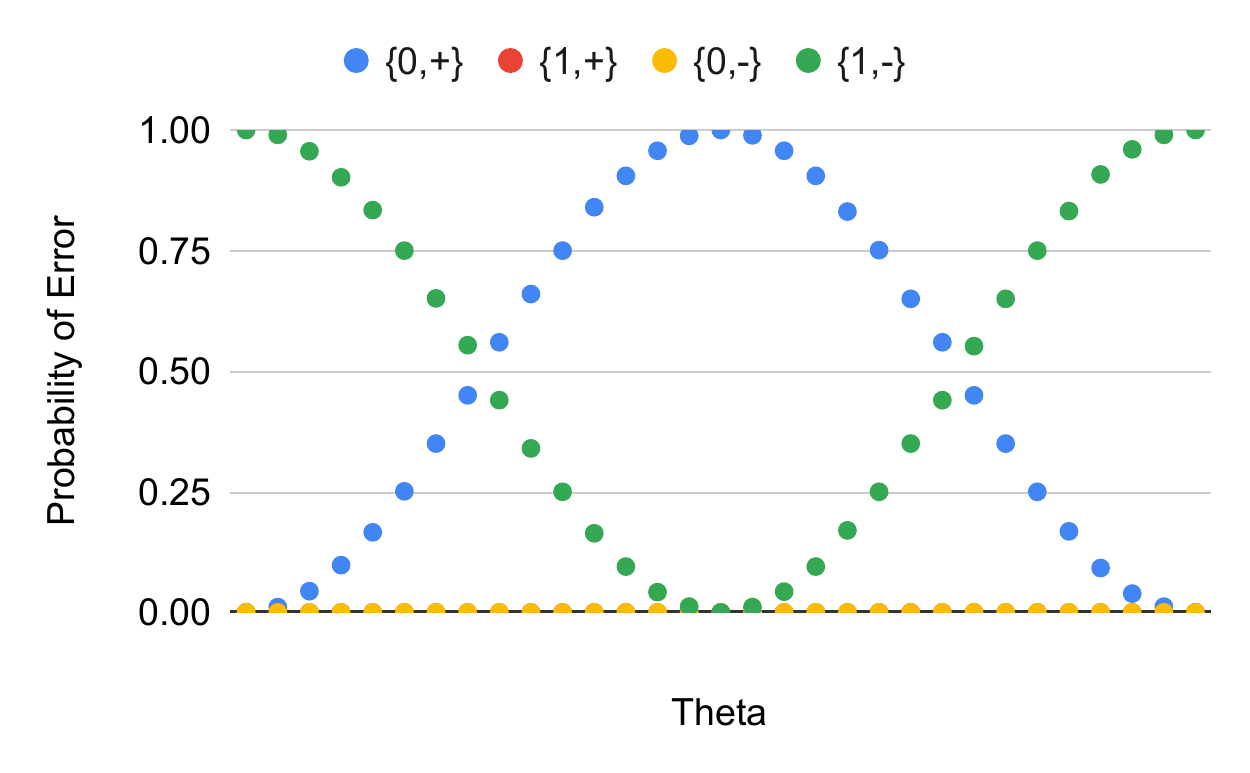}
\caption{\textbf{Probability of different types of errors for different Y-error magnitudes.} Probability of different types of errors is extracted from the simulation results of the 13-qubit entangled state with $\varepsilon=Y_{\theta}$ applied to first qubit (see Section \ref{qed_section3}) for different values of $\theta$ with $\theta\in[-\pi\ ,\pi]$. Here $\{0,+\},\{1,+\},\{0,-\}$ and $\{1,-\}$ represent the two qubit states $\Ket{00},\Ket{10},\Ket{01}$ and $\Ket{11}$ respectively, where $\Ket{+}=\frac{1}{\sqrt{2}}\big(\Ket{0}+\Ket{1}\big)$ and  $\Ket{-}=\frac{1}{\sqrt{2}}\big(\Ket{0}-\Ket{1}\big)$. $\Ket{+},\Ket{-}$ are the states of the second ancillary syndrome qubit before the Hadamard operation in the circuit of Fig. \ref{qed_fig2} in the bit-flip and phase-flip cases respectively. The blue line represents probability of no-error, the green line represents the probability of bit-flip as well as phase-flip error while the orange and yellow line represents probability of bit-flip and phase-flip errors respectively. We observe non vanishing error probability for both bit-flip and phase-flip errors as $Y_{\theta}$ can be decomposed as combination of bit-flip and phase-flip errors. Probability of no error shows a cosine dependence on $\theta$ which is expected since the matrix for $Y_{\theta}$ is given as $Y_{\theta}=\cos{(\theta/2)}I-i\sin{(\theta/2)}\sigma_y$ where $I$ is $2\times 2$ identity and $\sigma_y$ is the Pauli $y$ matrix.}

\label{qed_fig2}
\end{figure*}
 \begin{figure*}
  \centering
  \includegraphics[width=0.65\linewidth]{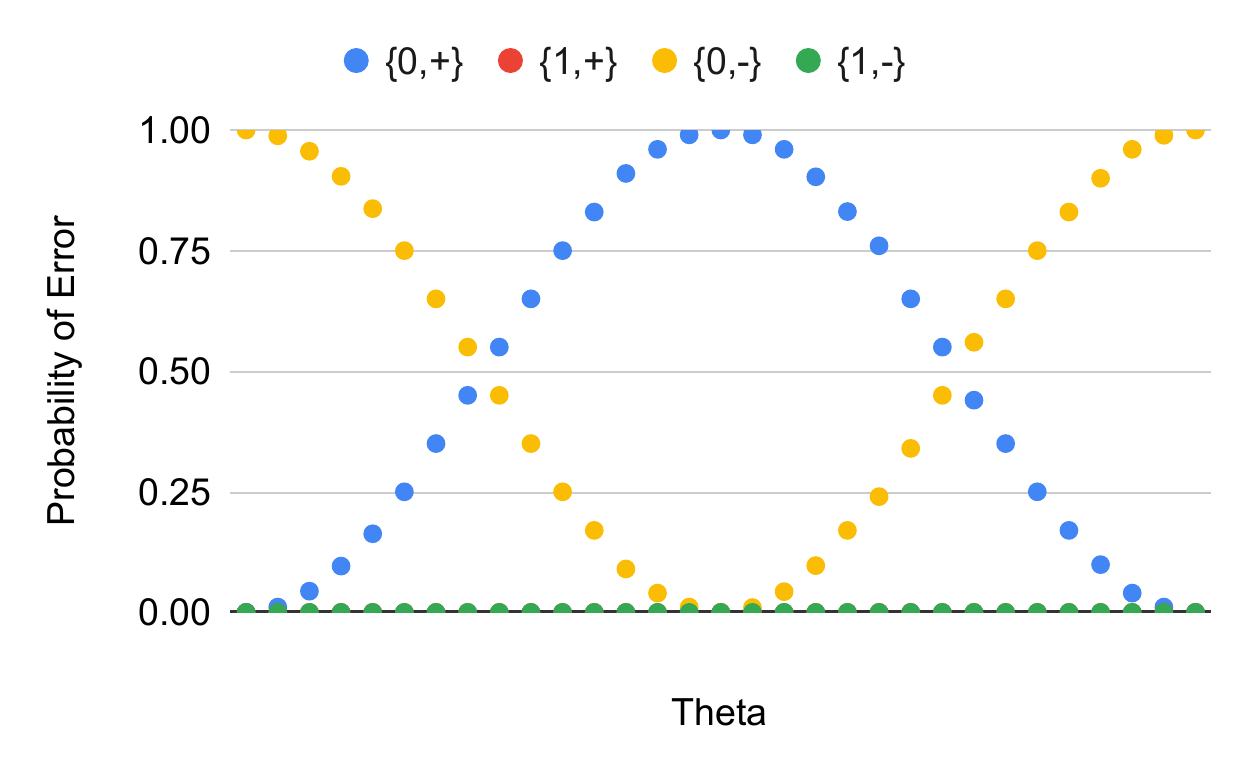}
  \caption{\normalsize{\textbf{Probability of different kinds of errors for arbitrary errors $X_{\theta}$.} Here $\{0,+\},\{1,+\},\{0,-\}$ and $\{1,-\}$ represent the two qubit states $\Ket{00},\Ket{10},\Ket{01}$ and $\Ket{11}$ respectively, where $\Ket{+}=\frac{1}{\sqrt{2}}\big(\Ket{0}+\Ket{1}\big)$ and  $\Ket{-}=\frac{1}{\sqrt{2}}\big(\Ket{0}-\Ket{1}\big)$. $\Ket{+},\Ket{-}$ are the states of the second syndrome qubit before the Hadamard operation in the circuit of Fig. \ref{qed_fig2} in the bit-flip and phase-flip cases respectively. The blue line represents probability of no-error, the green line represents the probability of bit-flip as well as phase-flip error while the orange and yellow line represents probability of bit-flip and phase-flip errors respectively. In this figure, non vanishing probability for bit-flip error and no-error is observed. This is because $X_{\theta}$ can be decomposed as $X_{\theta}=\cos{(\theta/2)}I-i\sin{(\theta/2)}\sigma_x$ where $\sigma_x$ is the Pauli $x$ matrix. The identity matrix in the decomposition is accounted for no-error and the Pauli $x$ matrix introduced bit-flip error.}}
  \label{qed_fig3a}
  \end{figure*} 
  \newpage
\begin{figure*}
  \centering 
\includegraphics[width=0.65\linewidth]{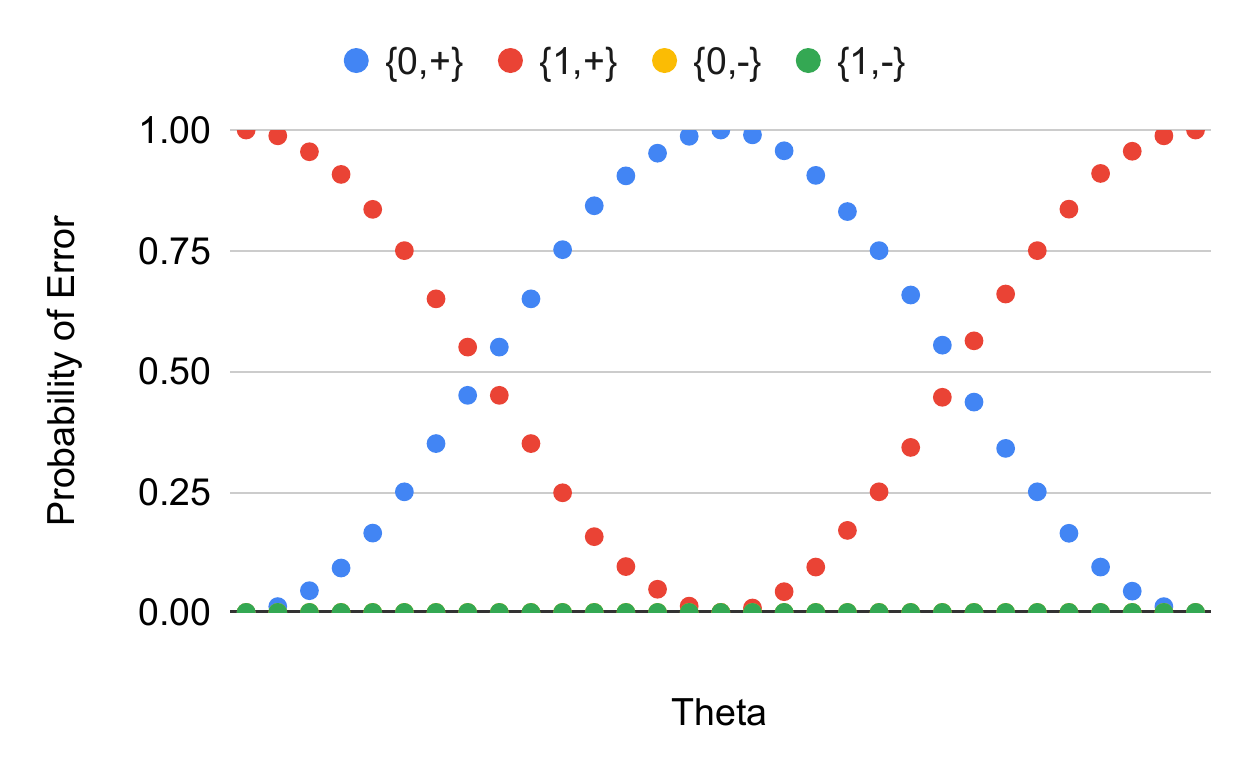}
  \caption{\normalsize{\textbf{Probability of different kinds of errors for arbitrary errors $Z_{\theta}.$} Here $\{0,+\},\{1,+\},\{0,-\}$ and $\{1,-\}$ represent the two qubit states $\Ket{00},\Ket{10},\Ket{01}$ and $\Ket{11}$ respectively, where $\Ket{+}=\frac{1}{\sqrt{2}}\big(\Ket{0}+\Ket{1}\big)$ and $\Ket{-}=\frac{1}{\sqrt{2}}\big(\Ket{0}-\Ket{1}\big)$. $\Ket{+},\Ket{-}$ are the states of the second syndrome qubit before the Hadamard operation in the circuit of Fig. \ref{qed_fig2} in the bit-flip and phase-flip cases respectively. The blue line represents probability of no-error, the green line represents the probability of bit-flip as well as phase-flip error while the orange and yellow line represents probability of bit-flip and phase-flip errors respectively. In this figure, non vanishing probability for phase-flip error and no-error is observed since again $Z_{\theta}$ can be decomposed as $Z_{\theta}=\cos{(\theta/2)}I-i\sin{(\theta/2)}\sigma_z$ where $\sigma_z$ is the Pauli $z$ matrix. The identity matrix in the decomposition is accounted for no-error and the Pauli $y$ matrix introduced phase-flip error.}}
  \label{qed_fig3b}
\end{figure*}
\subsection{Detecting arbitrary errors}
\begin{figure*}
\centering
\includegraphics[width=0.65\linewidth]{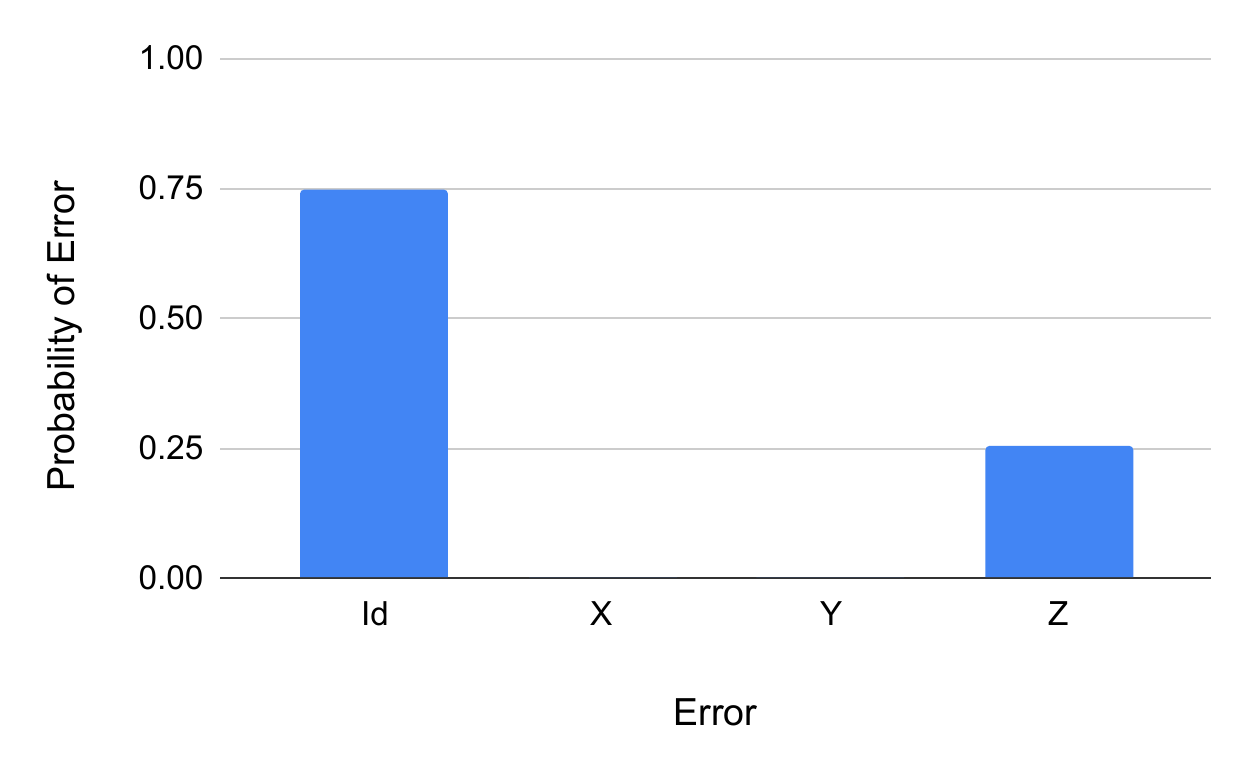}
\caption{\textbf{Detection of arbitrary errors.} The probability for each kind of error, no-error (Id), bit-flip ($X$), phase-flip ($Y$) and both bit-flip and phase-flip ($Z=XY$) is extracted from the measurement results of the syndrome qubits for the applied error $\varepsilon=Y_{\pi/3}$. We observe non zero probability for no error and bit-flip and phase-flip error simultaneously. This is because $Y_{\pi/3}$ can be decomposed as $Y_{\pi/3}=\frac{\sqrt{3}}{2}I-\frac{i}{2}\sigma_y$.}
\label{qed_fig4a}
\end{figure*}

\begin{figure*}
\centering
\includegraphics[width=0.65\linewidth]{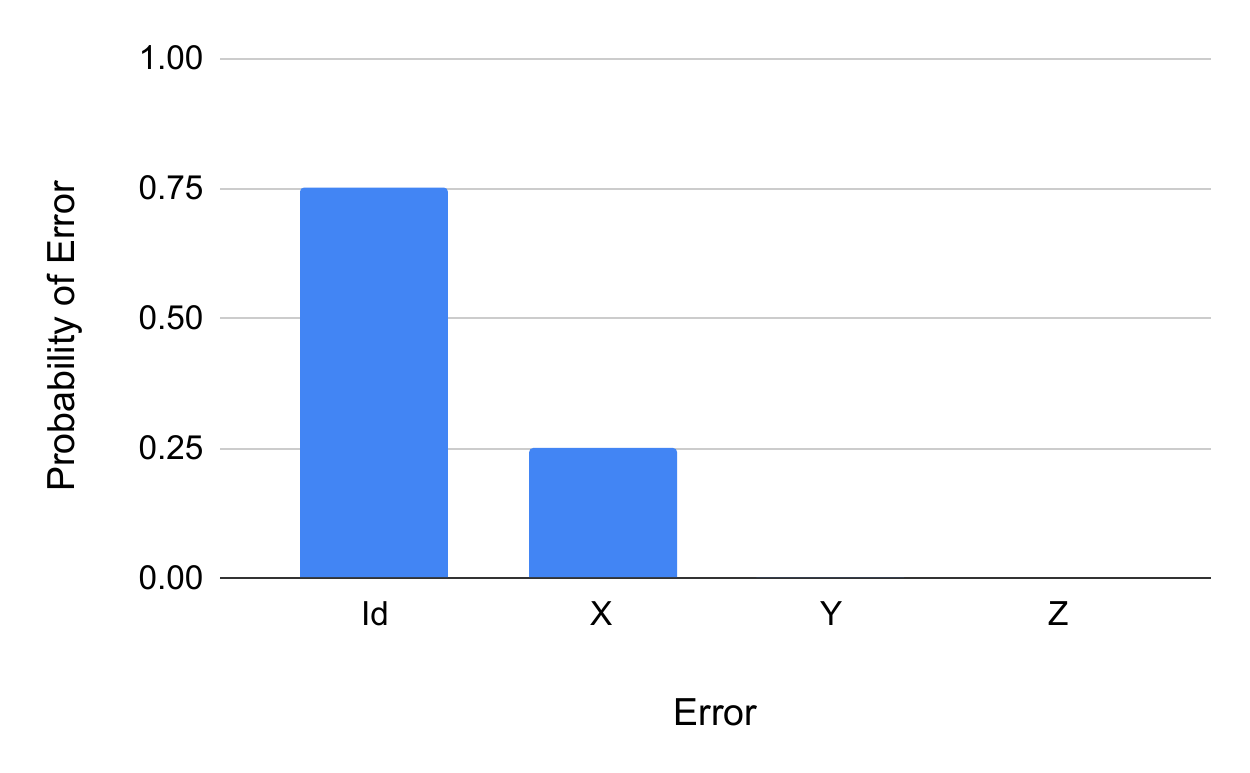}
\caption{\textbf{Detection of arbitrary errors.} The probability for each kind of error, no-error (Id), bit-flip ($X$), phase-flip ($Y$) and both bit-flip and phase-flip ($Z=XY$) is extracted from the measurement results of the syndrome qubits for the applied error $\varepsilon=X_{\pi/3}$. We observe non zero probability for no-error and bit-flip error because $X_{\pi/3}$ can be decomposed as $X_{\pi/3}=\frac{\sqrt{3}}{2}I-\frac{i}{2}\sigma_x$.}
\label{qed_fig4b}
\end{figure*}

\begin{figure*}
 \centering
\includegraphics[width=0.65\linewidth]{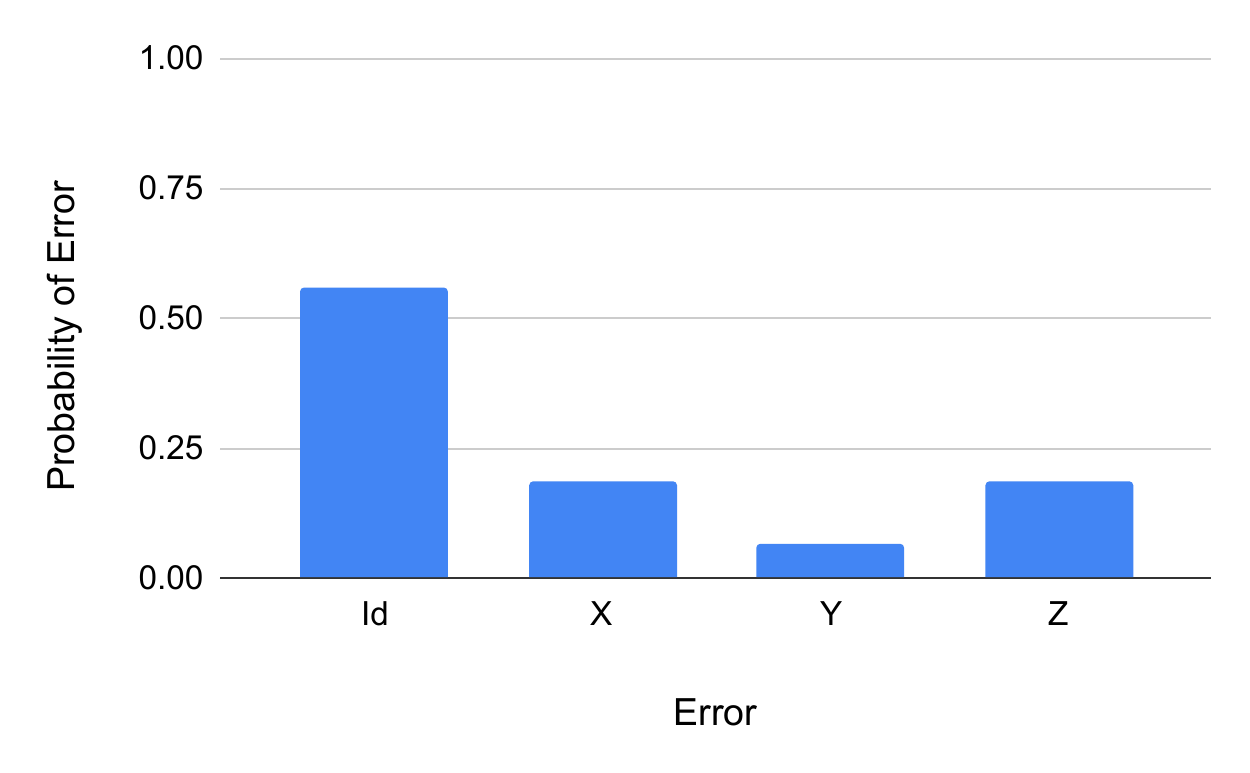}
\caption{\textbf{Detection of arbitrary errors.} The probability for each kind of error, no-error (Id), bit-flip ($X$), phase-flip ($Y$) and both bit-flip and phase-flip ($Z=XY$) is extracted from the measurement results of the syndrome qubits for the applied error $\varepsilon=X_{\pi/3}Y_{\pi/3}$. We observe non-zero probability for all types of error with different amplitudes as $X_{\theta}$ induces bit-flip error and $Y_{\theta}$ induces both bit-flip and phase-flip simultaneously.}
\label{qed_fig4c}
\end{figure*}

\begin{figure*}
\centering
\includegraphics[width=0.65\linewidth]{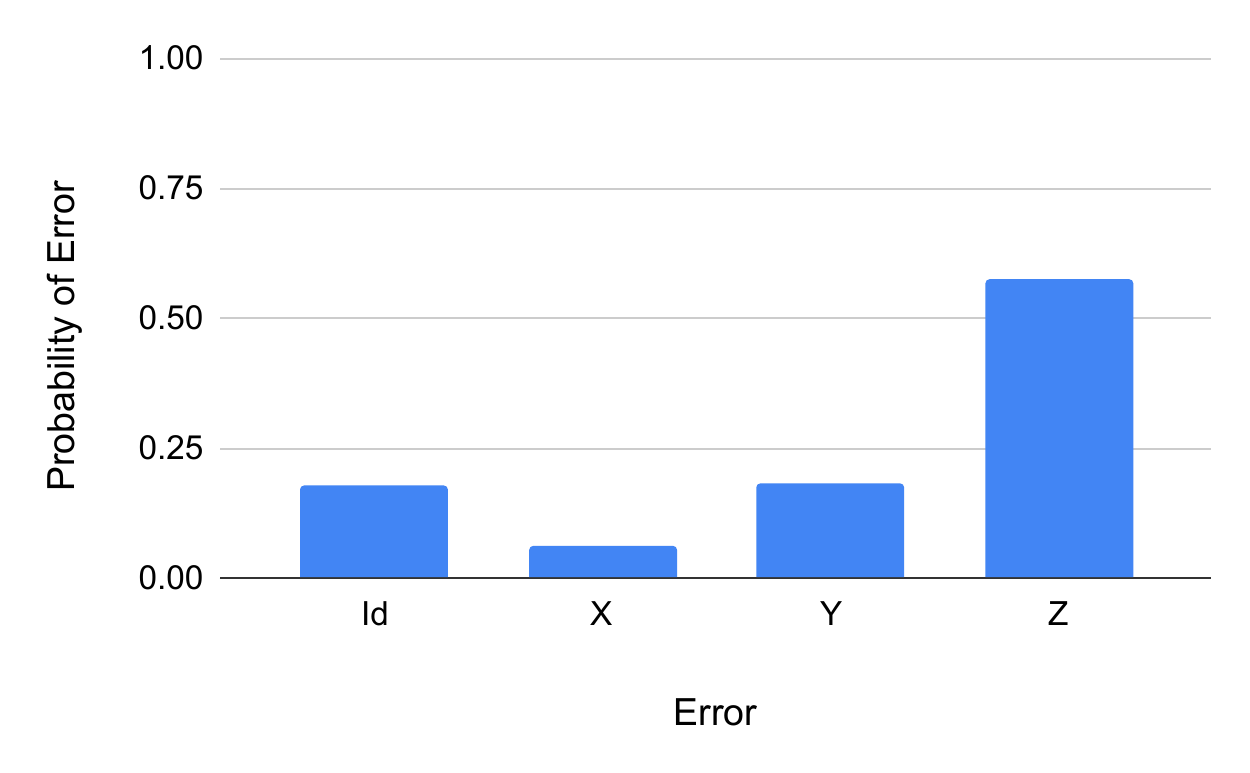}
\caption{\textbf{Detection of arbitrary errors.} The probability for each kind of error, no-error (Id), bit-flip ($X$), phase-flip ($Y$) and both bit-flip and phase-flip ($Z=XY$) is extracted from the measurement results of the syndrome qubits for the applied error $\varepsilon=X_{\pi/3}Y_{2\pi/3}$. We observe non-zero probability for all types of error with different amplitudes as $X_{\theta}$ induces bit-flip error and $Y_{\theta}$ induces both bit-flip and phase-flip simultaneously.}
\label{qed_fig4d}
\end{figure*}

\begin{figure*}
\centering
\includegraphics[width=0.65\linewidth]{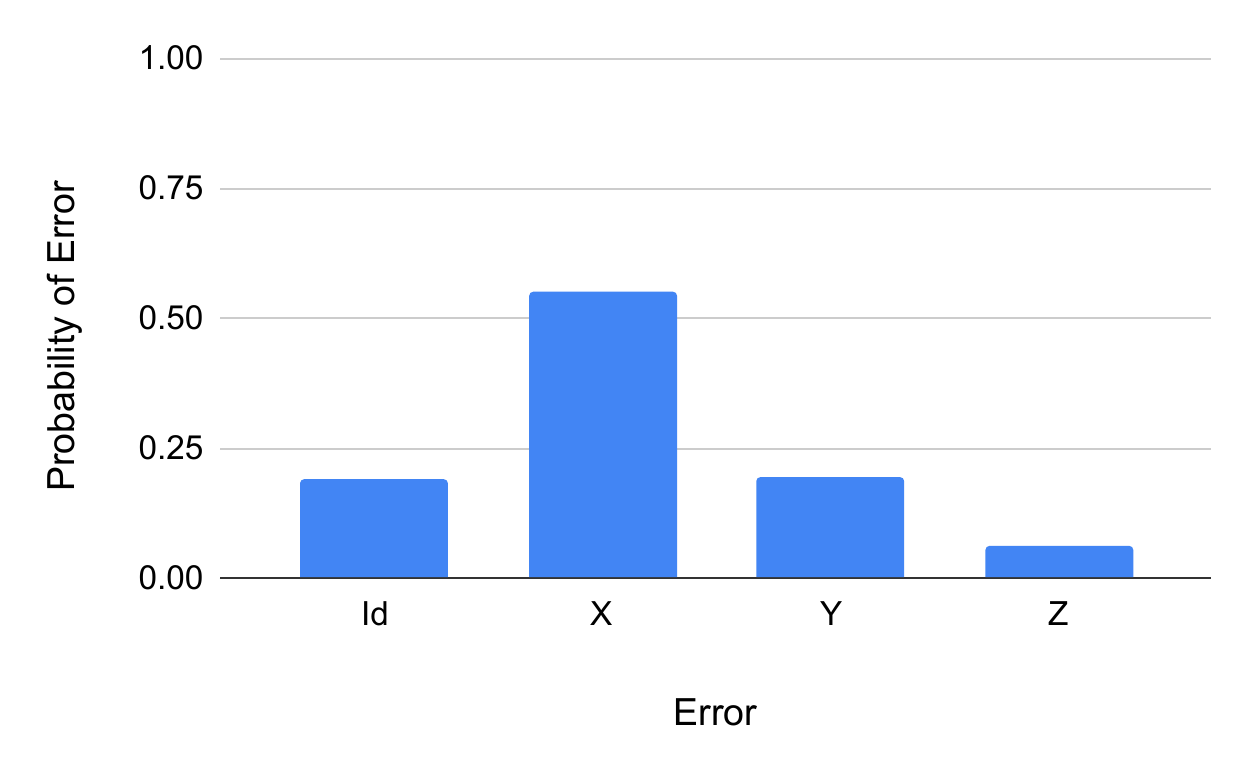}
\caption{\textbf{Detection of arbitrary errors.} For applied error $\varepsilon=X_{2\pi/3}Y_{\pi/3}$, we observe non zero probability for each type of error due to reasons mentioned in Fig. \ref{qed_fig4c} and Fig. \ref{qed_fig4d}.}
\label{qed_fig5a}
\end{figure*}

\begin{figure*}
\centering
\includegraphics[width=0.65\linewidth]{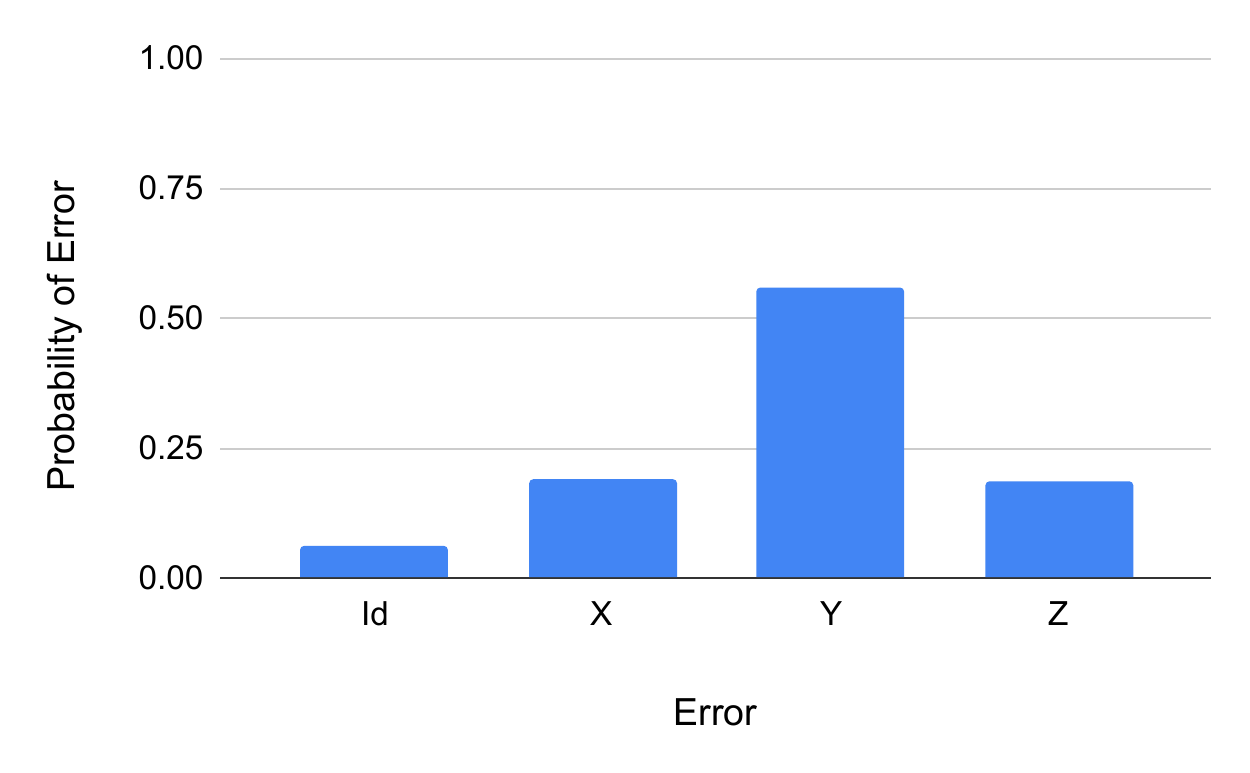}
\caption{\textbf{Detection of arbitrary errors.} For applied error $\varepsilon=X_{2\pi/3}Y_{2\pi/3}$, we observe non zero probability for each type of error due to reasons mentioned in Fig. \ref{qed_fig4c} and Fig. \ref{qed_fig4d}.}
\label{qed_fig5b}
\end{figure*}

\begin{figure*}
\centering
\includegraphics[width=0.65\linewidth]{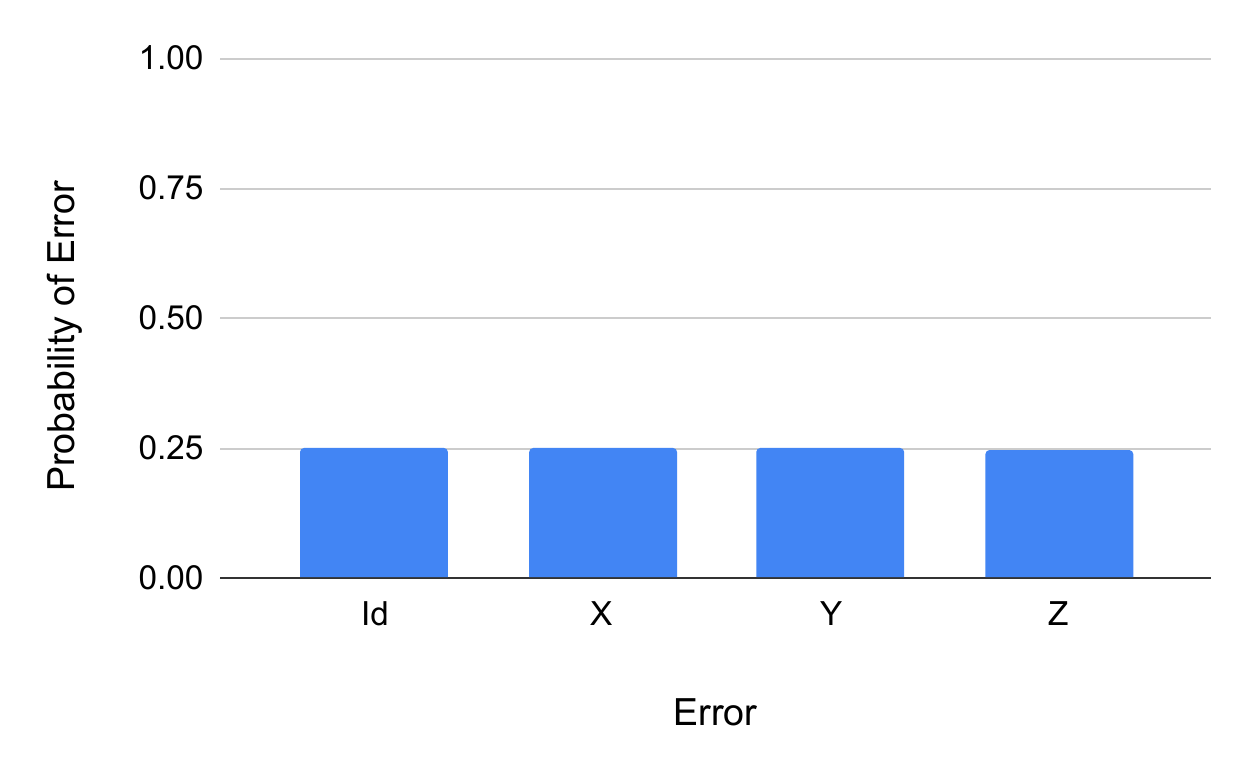}
\caption{\textbf{Detection of arbitrary errors.} For applied error $\varepsilon=R$, correspond to a $X_{\pi/2}Y_{\pi/2}$ operation, which maps the $x-y-z$ axes in the Block sphere to $y-z-x$, we observe almost equal probability for each type of error.}
\label{qed_fig5c}
\end{figure*}

\begin{figure*}
\centering
\includegraphics[width=0.65\linewidth]{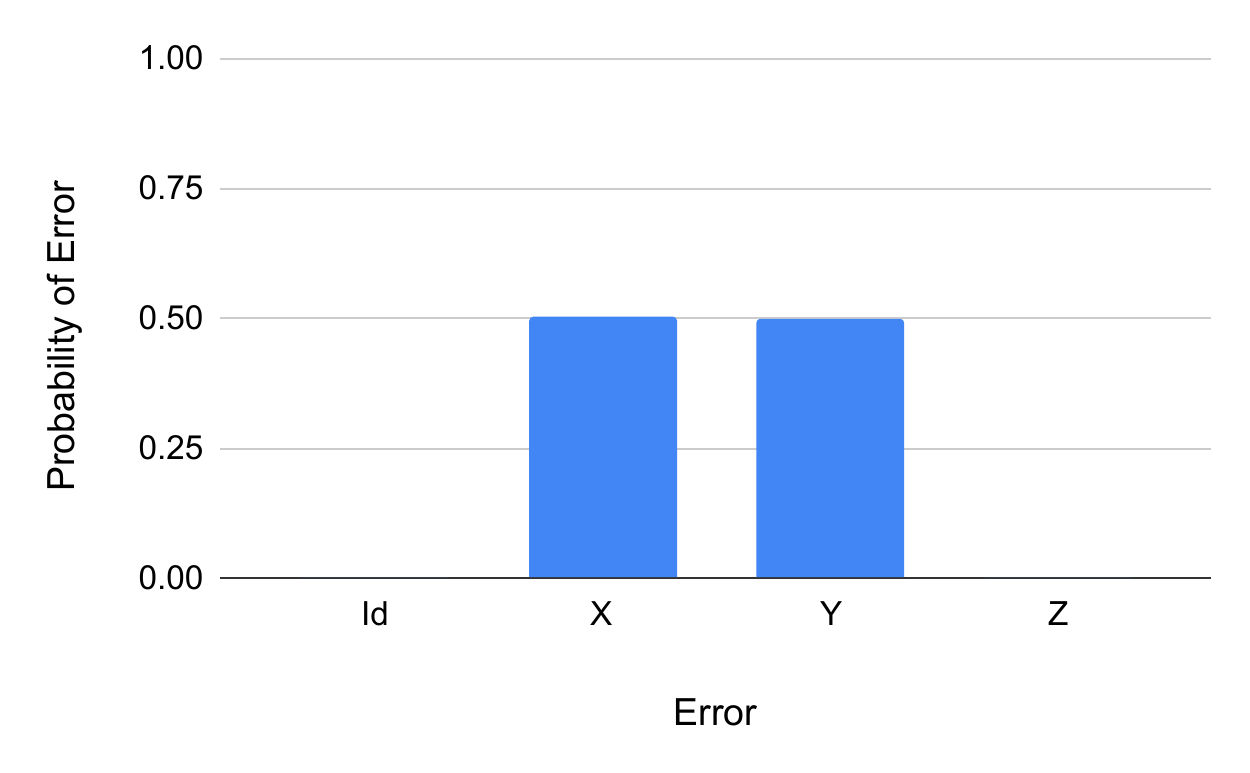}
\caption{\textbf{Detection of arbitrary errors.} For applied error $\varepsilon=H$ which correspond to the Hadamard operation, we observe equal probability for bit-flip and phase-flip error.}
\label{qed_fig5d}
\end{figure*}

Apart from single qubit bit-flip, phase-flip and both, the circuit in Fig. \ref{qed_fig1} also detects any arbitrary single qubit errors. The circuit presented here detects any arbitrary error in any single qubit from qubit 1 to qubit $2n$. The measurement result of the syndrome qubits can be tracked as $\theta$ is varied slowly between $-\pi$ and $\pi$ in an applied error $\varepsilon=Y_{\theta}$. In a 8192 shots simulation of the circuit in Fig. \ref{qed_fig2}, the probability of different errors is plotted as a function of $\theta$. The probability of different types of errors is compared (see Fig. \ref{qed_fig3a} and \ref{qed_fig3b}) for arbitrary errors $\varepsilon=X_{\theta}$ and $\varepsilon=Z_{\theta}$ varying the values of $\theta$ between $-\pi$ and $\pi$. 
To demonstrate arbitrary error detection, the error $\varepsilon$ is constructed via combinations of $X$ and $Y$ rotations. A set of eight arbitrary error used in the simulation are $\{Y_{\pi/3}, X_{\pi/3}, X_{\pi/3}Y_{\pi/3}, X_{\pi/3}Y_{2\pi/3}, X_{2\pi/3}Y_{\pi/3},\\ X_{2\pi/3}Y_{2\pi/3}, R, H\}$ where $R$ is $Y_{\pi/2}X_{\pi/2}$ and $H$ is the Hadamard operation. These errors are introduced in the first qubit of the 13-qubit entangled state and the probabilities of different types of errors is compared on a histogram for each of the eight errors (see Fig. \ref{qed_fig4a} - \ref{qed_fig5d}).
\newpage
\section{Methods}
\label{qed_section3}
\subsection{Initial entangled state}
In this subsection, we discuss the states with complimentarity property. We first show that the $2n$-qubit state of Eq. \eqref{qed_Eq.1} is entangled when $B_n$ is a nonempty proper subset of $A_n$ with the complimentarity property.
\begin{thm}
Let $B_n$ be a nonempty proper subset of $A_n$ with the complimentarity property. Then the state given by Eq. \eqref{qed_Eq.1} is entangled.
\label{thm III.1}
\begin{proof}
It suffices to prove that there exists a $|a_1a_2\dots a_i\dots a_{2n}\rangle\in B_n$ such that $|a_1a_2\dots (a_i\oplus 1)\dots a_{2n}\rangle\not\in B_n$ for all $1\leq i\leq 2n.$ We prove this by contradiction. Suppose there is no element in $B_n$ with the above property, then we prove that $A_n=B_n$ which is a contradiction. Let $|b_1b_2\dots b_i\dots b_{2n}\rangle\in A_n$ be an arbitrary element. By complementarity property of $|\psi\rangle_{B_n}$, it is clear that a basis vector of the form $|b_1c_2\dots c_{2n}\rangle\in B_n.$ Let $2\leq i\leq 2n$ be the smallest integer such that $c_i\neq b_i$. Then by assumption there is a basis vector of the form $|b_1b_2\dots b_{i-1} (c_i\oplus 1)c_{i+1}\dots c_{2n}\rangle\in B_n.$ Again let $i+1\leq j\leq 2n$ be the smallest integer such that $c_j\neq b_j$. Arguing as above, we see that a basis vector of the form $|b_1b_2\dots b_{i-1} b_ib_{i+1}\dots b_{j-1}(c_j\oplus 1)\dots c_{2n}\rangle\in B_n.$ Proceeding in this way, we see that $|b_1b_2\dots b_i\dots b_{2n}\rangle\in B_n\subset A_n$. This implies that $A_n=B_n$ and we are done.    
\end{proof}
\end{thm}
We now discuss graph states which form examples of states with complimentarity property when $B_n=A_n$. To each undirected graph $G=G(V,E)$ with $V$ the set of vertices and $E$ the set of edges, we associate an $V$-qubit quantum state defined as follows:
\[
|\psi\rangle_G=\frac{1}{2^{|V|/2}}\prod_{(a,b)\in E}U_{ab}\Ket{+}^{\otimes |V|},
\]
where $U_{ab}$ is the controlled-$Z$ gate acting on qubits $a$ and $b$ which correspond to the vertices $a$ and $b$ connected by the edge $(a,b)$ and \[\Ket{+}=\frac{1}{\sqrt{2}}(\Ket{0}+\Ket{1}).\]
It is clear that $\Ket{\psi}_G$ is a superposition of all the $2^{|V|}$ basis vectors and hence it trivially has the complementarity property. These states can be prepared using the following steps \cite{qed_WangarXiv:1801.037822018}:
\begin{enumerate}
\item Initialize the state to $|+\rangle^{\otimes |V|}$ by applying $|V|$ Hadamard gates to $|0\rangle^{\otimes |V|}.$
\item For every $(a, b) \in E$, apply a control-$Z$ gate on qubits $a$ and $b$; the order can be arbitrary.
\end{enumerate}
It is a nontrivial result that for any general undirected graph $G$, $\Ket{\psi}_G$ is an entangled state \cite{graphstateent}. The particular graph state that we use to simulate our error-detection protocol is a $12$-qubit graph state corresponding to the 12 vertex ring graph. The circuit to prepare this state is shown in the blue box in Fig. \ref{qed_fig7} and is taken from \cite{qed_WangarXiv:1801.037822018}.

\begin{thm}
The state obtained after adding another qubit $\Ket{0}_{2n+1}$ to the state in Eq. \eqref{qed_Eq.1} using the circuit in Fig. \ref{qed_fig1} (blue, smaller yellow and red box) is entangled if and only if there exists a basis vector in the sum of Eq. \eqref{qed_Eq.1} with odd number of 1s in it and another basis vector with even number of 1s in it.
\label{thm III.2}
\begin{proof}
First suppose that the sum in Eq. \ref{qed_Eq.1} has a basis vector with odd number of 1s and another basis vector with even number of 1s. Denote by $B_n^{\text{odd}}\subset B_n$ be the set of basis vectors with odd number of 1s and $B_n^{\text{even}}\subset B_n$ be the set of basis vectors with even number of 1s. Then we can write the state $|\psi\rangle_{B_n}$ as
\[
|\psi\rangle_{B_n}=\frac{1}{\sqrt{|B_n|}}\left[\sum_{\textbf{x}\in B_n^{\text{odd}}}\pm|\textbf{x}\rangle +\sum_{\textbf{y}\in B_n^{\text{even}}}\pm|\textbf{y}\rangle\right].
\]
It is easy to see that after all the $CNOT$ operations on $\Ket{0}$ in the circuit in Fig. \ref{qed_fig1}, it remains $\Ket{0}$ with the terms containing even number of $1s$ and becomes $\Ket{1}$ with the terms containing odd number of $1s$. 
Thus the new $(2n+1)$-qubit state has the form 
\[
|\psi\rangle_{2n+1}=\frac{1}{\sqrt{|B_n|}}\left[\sum_{\textbf{x}\in B_n^{\text{odd}}}\pm|\textbf{x}\rangle|1\rangle +\sum_{\textbf{y}\in B_n^{\text{even}}}\pm|\textbf{y}\rangle|0\rangle\right].
\]
Thus the state obtained after adding $\Ket{0}$ cannot be factored. Thus the entangled state in Eq. \ref{qed_Eq.1} remains entangled after addition of the qubit $\Ket{0}$. Conversely, suppose that the sum in Eq. \ref{qed_Eq.1} contains only basis vectors with odd or even number of $1s$ in it but not both. After all the CNOT operations on $\Ket{0}$ as shown in the smaller yellow and red box in Fig. \ref{qed_fig1}, it remains $\Ket{0}$ for the case when only terms with even number of $1s$ are there in the sum and the new $(2n+1)$-qubit state is $|\psi\rangle_{B_n}\Ket{0}$ which is not entangled. Similarly, when only terms with odd number of terms are there in the sum of Eq. \eqref{qed_Eq.1}, the new $(2n+1)$-qubit state is $|\psi\rangle_{B_n}\Ket{1}$ which is again not entangled. Thus we have proved the contrapositive of the converse part of the theorem.
\end{proof}
\end{thm}
\subsection{Analysis of the Protocol}
Here, we verify the measurement results listed in Table \ref{qed_table1}. Consider a $2n$-qubit state given in
Eq. \eqref{qed_Eq.1}. We split the analysis into two cases :
\begin{enumerate}
\item \textbf{The $2n$-qubit state contains terms with only odd or even number of $1$s :}
Arguing as in the proof of the converse part of Theorem \ref{thm III.2}, the $(2n+1)$-qubit state $\Ket{\psi}_{2n+1}$ has the form
\[
\Ket{\psi}_{2n+1}=\Ket{\psi}_{B_n}\Ket{0}\quad\text{or}\quad\Ket{\psi}_{B_n}\Ket{1}
\]
depending on whether the terms in $\Ket{\psi}_{B_n}$ has even or odd number of 1s. But now it can be easily seen that every term in $\Ket{\psi}_{2n+1}$ contains even number of 1s in both the cases. It can be observed easily from the circuit that the first syndrome qubit always remains in the state $\Ket{0}$ and hence the state with first syndrome qubit has the form 
\[
\Ket{\psi}_{2n+2}=\Ket{\psi}_{B_n}\Ket{0}\Ket{0}_{s_1}\quad\text{or}\quad \Ket{\psi}_{B_n}\Ket{1}\Ket{0}_{s_1},
\]
where $s_1$ denotes the first syndrome qubit.
The second syndrome qubit $s_2$
acts as the control in the CNOT operation applied on the first $2n$-qubits as shown in Fig. \ref{qed_fig1}. After the first Hadamard operation on $s_2$, the $\Ket{0}$ in $\Ket{+}$ leaves the first
$2n$-qubits unchanged while the $\Ket{1}$ changes each term in $\Ket{\psi}_{B_n}$ state to its complementary term. Thus we see that after the second Hadamard operation on $s_2$ and just before measurement on the syndrome qubits, the final $2n+3$ qubit state has the form
\[
\begin{split}
\Ket{\psi}_{2n+3}=\Ket{\psi}_{B_n}\Ket{0}\Ket{0}_{s_1}\Ket{0}_{s_2}\quad\text{or}\\\Ket{\psi}_{2n+3}= \Ket{\psi}_{B_n}\Ket{1}\Ket{0}_{s_1}\Ket{0}_{s_2},
\end{split}
\]
\begin{itemize}
\item \textbf{No error:} When we introduce no error in any of the qubits, the measurement result gives $\Ket{00}$ as is evident from the form of $\Ket{\psi}_{2n+3}$.
\item \textbf{Bit-flip error:} If we introduce a bit-flip error in any one of the $2n+1$ qubits, then each term in $\Ket{\psi}_{2n+1}$ has odd number of 1s and thus after the CNOT operation on $s_1$ as shown in Fig. \ref{qed_fig1}, it changes to $\Ket{1}$. Also the second syndrome qubit remains intact since bit-flip does not destroy the complementarity property of $\Ket{\psi}_{B_n}$. The $2n+3$ qubit state before is then given by
\[
\begin{split}
    \Ket{\psi}_{2n+3}=\Ket{\psi}_{B_n}^{\varepsilon}\Ket{0}\Ket{1}_{s_1}\Ket{0}_{s_2}\quad\text{or}\\\Ket{\psi}_{2n+3}=\Ket{\psi}_{B_n}^{\varepsilon}\Ket{1}\Ket{1}_{s_1}\Ket{0}_{s_2},
\end{split}
\]
where $\Ket{\psi}_{B_n}^{\varepsilon}$ is the state $\Ket{\psi}_{B_n}$ with error. Hence, the measurement result turns out to be $\Ket{10}$.
\item \textbf{Phase-flip error :} Now, suppose we introduce a phase-flip error in the $i^{\text{th}}$ qubit, $1\leq i\leq 2n$. Observe that we can write $\Ket{\psi}_{B_n}$ as
\[
\Ket{\psi}_{B_n}=\Ket{\psi}_{2n-1}\Ket{0}_i+\Ket{\psi}_{2n-1}^c\Ket{1}_i,
\]
where $\Ket{\psi}_{2n-1}$ is some $2n-1$ qubit state and $\Ket{\psi}_{2n-1}$ is the same state with its terms being the complimentary of the terms of $\Ket{\psi}_{2n-1}$. After the phase-flip error is introduced, the state becomes
\[
\Ket{\psi}_{B_n}^{\varepsilon}=\Ket{\psi}_{2n-1}\Ket{0}_i-\Ket{\psi}_{2n-1}^c\Ket{1}_i.
\]
Since this error does not destroy the complimentarity property, we see that the $2n+2$ qubit state is given by 
\[
\begin{split}
    &\Ket{\psi}_{2n+2}=(\Ket{\psi}_{2n-1}\Ket{0}_i-\Ket{\psi}_{2n-1}^c\Ket{1}_i)\Ket{0}\Ket{0}_{s_1}\\ &\quad\quad\quad\text{or}\\&\Ket{\psi}_{2n+2}=(\Ket{\psi}_{2n-1}\Ket{0}_i-\Ket{\psi}_{2n-1}^c\Ket{1}_i)\Ket{1}\Ket{0}_{s_1}.
\end{split}
\]
Finally $s_2$ acts as control for the CNOT operations on the first $2n$ qubits as shown in the circuit in Fig. \ref{qed_fig1}. The state just before the last Hadamard gate is then given by
\[
\begin{split}
    \Ket{\psi}_{2n+3}&=\frac{(\Ket{\psi}_{2n-1}\Ket{0}_i-\Ket{\psi}_{2n-1}^c\Ket{1}_i)\Ket{0}\Ket{0}_{s_1}\Ket{0}_{s_2}}{\sqrt{2}}\\&+\frac{(\Ket{\psi}_{2n-1}^c\Ket{1}_i-\Ket{\psi}_{2n-1}\Ket{0}_i)\Ket{1}\Ket{0}_{s_1}\Ket{1}_{s_2}}{\sqrt{2}}\\&=(\Ket{\psi}_{2n-1}\Ket{0}_i-\Ket{\psi}_{2n-1}^c\Ket{1}_i)\Ket{0}\Ket{0}_{s_1}\Ket{-}_{s_2}.
\end{split}
\]
Hence, it is clear that the measurement result will be $\Ket{01}$.
\item \textbf{Bit-flip and Phase-flip error}
With similar analysis, it is easy to see that if both the errors are introduced in one of the $2n$-qubits then the two syndrome qubits have the state $\Ket{1}_{s_1}$ and $\Ket{-}_{s_2}$ respectively just before the last Hadamard operation and thus the measurement result is $\Ket{11}$. 
\end{itemize}
\item \textbf{The $2n$-qubit state contains a ket with odd number of 1s and another ket with even number of 1s :} In this case the state obtained after addition of $\Ket{0}$ to the $2n$-qubit state is entangled and every term in this $(2n+1)$-qubit entangled state contains even number of 1s. Thus following the same analysis as done in the above case gives us the same result.
\end{enumerate}
It is interesting to note that all the maximally entangled Bell states and GHZ state with even number of qubits fall under the first case. Thus using the protocol, any single-qubit phase-change error or phase-flip or bit-flip error can be detected. A limitation of the protocol is that phase-flip error on the $(2n+1)^{\text{th}}$-qubit could not be detected. To understand this we first write our $(2n+1)$-qubit state as
\begin{equation}
|\psi\rangle=\left|\psi_{1}\right\rangle|0\rangle+\left|\psi_{2}\right\rangle|1\rangle
\label{eq 4}
\end{equation}
where the normalization constant is absorbed in $\left|\psi_{1}\right\rangle$ and $\left|\psi_{1}\right\rangle$ which are $2 n$-qubit states. An easy observation shows that $\left|\psi_{1}\right\rangle$ and $\left|\psi_{2}\right\rangle$ have the complementarity property independently. Now a phase-flip error in the last qubit changes the above state to
\[
|\psi\rangle=\left|\psi_{1}\right\rangle|0\rangle-\left|\psi_{2}\right\rangle|1\rangle
\]
After applying all the CNOT operations given in the circuit in Fig. \ref{qed_fig1}, we note that from both the terms
in Eq. \eqref{eq 4}, the two syndrome qubits factor out as $|0\rangle$ and $\frac{1}{\sqrt{2}}(|0\rangle+|1\rangle)$ leaving the state in Eq. \eqref{eq 4} unchanged. Thus the protocol fails to detect a phase-flip error in the $(2n+1)^{\text{th}}$-qubit.
%\clearpage

\subsection{Simulation of the protocol on the 16-qubit IBM quantum computer.}
We simulate the protocol for a 13-qubit entangled state on 16-qubit \textit{ibmqx5} IBM quantum computer. \textit{ibmqx5} is a 16-qubit quantum processor made up of superconducting qubits based on transmon qubits \cite{qed_KochPRA2007}.

\begin{comment}
These qubits are not susceptible to charge noise. Thus these provide an efficient quantum computing experience. Since the advent of real quantum computing in 2016 by IBM, it provides a cloud-based access to its experimental quantum computing platform called IBM Q Experience \cite{qed_IBM}. IBM provides a cloud based access to a 5-qubit device called IBM Q5 or \textit{ibmqx4} quantum computer and a 16-qubit device called IBM Q16 or \textit{ibmqx5} quantum computer. In our simulation, we use the 16-qubit \textit{ibmqx5} quantum computer. The details of various parameters of each of the 16 qubits of the \textit{ibmqx5} quantum computer is listed in table \ref{qed_table2} and table \ref{qed_table3} \cite{qed_QISKit}. The architecture of the 16-qubit quantum computer is shown in Fig. \ref{qed_fig6} \cite{qed_IBM}. 
\end{comment}

\begin{figure*}[]
\centering
\includegraphics[width=\linewidth]{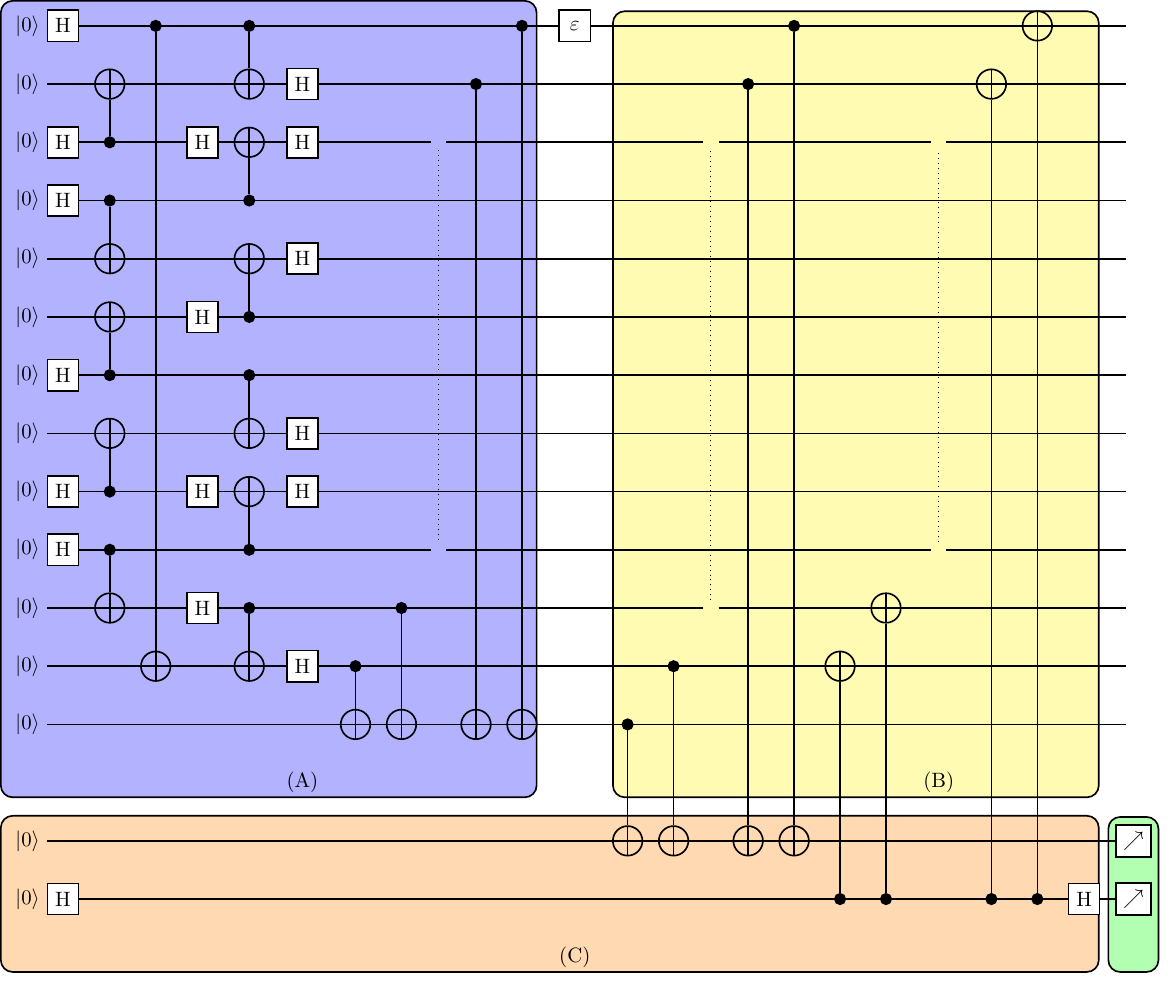}
\caption{\textbf{Circuit for simulation of error detection protocol in IBM Q16 quantum computer.} In the circuit, the first 12 qubits in box A are the 12-qubit entangled state with complementarity property. The 13th qubit is added to generate a 13-qubit entangled state. Qubits in box A is the initial state on which the error will be detected. The two qubits in box C represent the two syndrome qubits. Box B depicts the connection of syndrome qubit with the initial entangled state. The box with $\varepsilon$ is the error to be introduced in the entangled state. The last box (green) represents the measurement operations on the syndrome qubits.}
\label{qed_fig7}
\end{figure*}

A web-based quantum circuit construction is provided by IBM for Q5 which is run by simulation or real experiment. To compose quantum circuits, QASM language is needed. These circuits can then be run via simulation or real experiment using QISKit Python SDK \cite{qed_IBM,qed_QISKit}. In our simulation, we first prepare a 12-qubit entangled state with the complementarity property. For this, we use the quantum circuit proposed by Yuanhao \textit{et al.} \cite{qed_WangarXiv:1801.037822018}. Then we entangle another qubit with this 12-qubit entangled state using the CNOT operations as shown in Fig. \ref{qed_fig7}. We write the QISKit code for our circuit and then run the simulation with 8192 shots and record the number of times each result in the measurement of the two error syndrome qubits occur. Using those numbers, we calculate the probability of each kind of error (For data see Supplementary Information). Bit-flip and phase-flip errors are introduced using $X$ gate and $U_1$ gate respectively. Both bit-flip and phase-flip error simultaneously are introduced by applying $X$ gate and $U_1$ gate simultaneously. Arbitrary phase change and axis rotation errors are introduced using $U_3$ gate. For example, the error $Y_{\theta}$ is introduced by the operation $U_3(\theta, 0, 0)$ whereas the errors $X_{\theta}$ and $Z_{\theta}$ are introduced using $U_3(\theta, \pi/2, -\pi/2)$ and $U_1(\theta)$ respectively. The recorded probabilities are then compared for different kinds of errors (details in Sec. \ref{qed_section2}).\\
\begin{comment}
\begin{figure}[!ht]
\centering
\includegraphics[width=1\linewidth]{qed_fig6.pdf}
\caption{\textbf{Architecture of IBM Q16 quantum processor.} The picture shows the clip layout of 16-qubit quantum processor Rueschlikon [$ibmqx5$]. The connectivity of $CNOT$ operations among the 16 qubits are depicted. Allowed $CNOT$ operations are $Q1\longrightarrow [Q0,Q2]$, $Q2\longrightarrow Q3$, $Q3\longrightarrow [Q14,Q4]$, $Q5\longrightarrow Q4$, $Q6\longrightarrow [Q11,Q7,Q5]$, $Q8\longrightarrow Q7$, $Q7\longrightarrow Q10$, $Q9\longrightarrow [Q8,Q10]$, $Q11\longrightarrow Q10$, $Q12\longrightarrow [Q13,Q11,Q5]$, $Q13\longrightarrow [Q14,Q4]$, $Q15\longrightarrow [Q14,Q2,Q0]$, where $Qi\longrightarrow Qj$ means $Qi$ is the control bit and $Qj$ is the target bit.}
\label{qed_fig6}
\end{figure}
\end{comment}
\section{Comparative Analysis of the Error Detection Code}
In this section, we compare and contrast our proposed protocol of error detection with the existing error detection codes. We prominantly compare the computational cost of our protocol with the results of Corcoles \textit{et. al.} \cite{qed_CorcolesNatComm2015} and Linke \textit{et. al.} \cite{qed_LinkeScienceAdvances2017}. From the analysis in previous section, we conclude that the code is able to detect an arbitrary quantum error in any one of the first $2n$ qubits of the $(2n+1)$-qubit entangled state and detects any bit-flip error on the last qubit of the $(2n+1)$-qubit entangled state via measurements on a pair of ancillary error syndrome qubits. Thus we only need two ancillary qubits and are able to detect errors in all of the qubits of the state. To compare this with the code of Corcoles \textit{et. al.}, we emphasize that their code requires two ancillary qubits and can detect error only on two qubit entangled state. On the other hand, the code proposed by Linke \textit{et. al.} detects errors in a four qubit state using one ancillary qubit but with two qubits acting as stabilizers. Hence, we see that our code is far more computationally superior than the existing codes.

\section{Discussion \label{qed_discussion}}
We have provided an error detection code for a $(2n+1)$-qubit entangled system, with a general form which encompasses many important entangled states such as maximally entangled Bell states and generalized GHZ states, and simulated its usefulness on IBM Q16 quantum computer. As an immediate consequence of the protocol we notice that any arbitrary single qubit phase change error or bit-flip and phase-flip error in GHZ states with even number of qubits can be detected using the code. We explicitly checked our code for a 13-qubit entangled state and concluded that the code works well and detects any arbitrary single qubit phase-change error or bit-flip or phase-flip error in any of the first 12 qubits and detects any bit-flip error on the 13th qubit. In conclusion, we have provided an error detection code which can be helpful in manipulating quantum algorithm. This code can be used by different research groups to generate error detection codes for their quantum chips. In future, the work can be extended to the code generalizing all kinds of entangled states although the entangled state we used is quite general.

\section*{Data availability}
Data are available to any reader upon reasonable request.

\section*{Acknowledgments}
\label{qlock_acknowledgments}
The authors would like to thank the anonymous referee for a detailed feedback whch led to the improvement of the paper. R.K.S acknowledges the financial support of Indian Academy of Sciences (IAS). R.K.S would also like to thank Indian Institute of Science Education and Research Kolkata for providing hospitality during the course of the project. B.K.B. acknowledges the support of Inspire Fellowship awarded by DST, Government of India. The authors acknowledge the support of IBM Quantum Experience for producing experimental results. The views expressed are those of the authors and do not reflect the official policy or position of IBM or the IBM Quantum Experience team.
\section*{Author contributions}
R.K.S. and B.K.B. have developed the quantum error detection code. R.K.S., B.P. and B.K.B. have discussed and designed all the quantum circuits, and have discussed the data and analysed the data.  R.K.S. and B.P. performed all the experiments in IBM Quantum Experience platform.  B.K.B. supervised the project. P.K.P. has thoroughly checked and reviewed the manuscript. R.K.S., B.P. and B.K.B. have completed the project under the guidance of P.K.P.  

\section*{Competing interests}
The authors declare no competing financial as well as non-financial interests.
\section{Supplementary Information: Demonstration of a general fault-tolerant quantum error detection code for $(2n+1)$-qubit entangled state on IBM 16-qubit quantum computer}
For simulating the error detection protocol, we used QISKit to take both simulation results. The QASM code for the same is as follows: 

\begin{lstlisting}[language=Python]
# Import the QISKit SDK
from qiskit import QuantumCircuit, ClassicalRegister, QuantumRegister
from qiskit import available_backends, execute
QX_TOKEN="9e333de2e3732e2f9b354939a8ea4336c2
0c87403c930a944ed3fd28112a40a594ca0a4165a9f0
1878c8aacdf97af6b42f237048b836a8f8277bfd7d3b
3b4663"
QX_URL = "https://quantumexperience.ng.bluemix.net/api"
q = QuantumRegister(15)
c = ClassicalRegister(15)

qc = QuantumCircuit(q, c)
from math import pi

# We are first prepare our entangled state

qc.h(q[0])
qc.h(q[2])
qc.h(q[3])
qc.h(q[6])
qc.h(q[8])
qc.h(q[9])
qc.cx(q[2], q[1])
qc.cx(q[3], q[4])
qc.cx(q[6], q[5])
qc.cx(q[8], q[7])
qc.cx(q[9], q[10])
qc.cx(q[0], q[11])

qc.h(q[2])
qc.h(q[5])
qc.h(q[8])
qc.h(q[10])

qc.cx(q[0], q[1])
qc.cx(q[3], q[2])
qc.cx(q[5], q[4])
qc.cx(q[6], q[7])
qc.cx(q[9], q[8])
qc.cx(q[10], q[11])
qc.h(q[1])
qc.h(q[2])
qc.h(q[4])
qc.h(q[7])
qc.h(q[8])
qc.h(q[11])

#  Addition of another qubit to make 13 qubit entangled state

qc.cx(q[0], q[12])
qc.cx(q[1], q[12])
qc.cx(q[2], q[12])
qc.cx(q[3], q[12])
qc.cx(q[4], q[12])
qc.cx(q[5], q[12])
qc.cx(q[6], q[12])
qc.cx(q[7], q[12])
qc.cx(q[8], q[12])
qc.cx(q[9], q[12])
qc.cx(q[10], q[12])
qc.cx(q[11], q[12])

# Introduction of Error


# Addition of first ancillary error syndrome qubit


qc.cx(q[0], q[13])
qc.cx(q[1], q[13])
qc.cx(q[2], q[13])
qc.cx(q[3], q[13])
qc.cx(q[4], q[13])
qc.cx(q[5], q[13])
qc.cx(q[6], q[13])
qc.cx(q[7], q[13])
qc.cx(q[8], q[13])
qc.cx(q[9], q[13])
qc.cx(q[10], q[13])
qc.cx(q[11], q[13])
qc.cx(q[12], q[13])

# Addition of second ancillary error syndrome qubit

qc.h(q[14])

qc.cx(q[14], q[0])
qc.cx(q[14], q[1])
qc.cx(q[14], q[2])
qc.cx(q[14], q[3])
qc.cx(q[14], q[4])
qc.cx(q[14], q[5])
qc.cx(q[14], q[6])
qc.cx(q[14], q[7])
qc.cx(q[14], q[8])
qc.cx(q[14], q[9])
qc.cx(q[14], q[10])
qc.cx(q[14], q[11])

qc.h(q[14])



qc.measure(q[13], c[13]);qc.measure(q[14], c[14])  
# See a list of available local simulators
print("Local backends: ", available_backends({'local': True}))

# Compile and run the Quantum circuit on a simulator backend
job_sim = execute(qc, "local_qasm_simulator", shots=8192, max_credits=10)
sim_result = job_sim.result()

# Show the results
print("simulation: ", sim_result)
print(sim_result.get_counts(qc))
\end{lstlisting}
\subsection*{Measurement data}
We performed all the simulations on QISKit and recorded the countings of each of the measurement result over the two ancillary error syndrome qubit in 8192 shots. From the countings, the probability of each error \textit{i.e.} bit-flip error, phase-flip error and arbitrary phase-change error was extracted. The data is shown in the table \ref{qed_table4} below.
\begin{table*}[h!]
\addtolength{\tabcolsep}{20pt}
 \begin{tabular}{|c c c c c|} 
 \hline
 \hline
 Error & $\{0 , +\}$ & $\{1 , +\}$&$\{0 , -\}$&$\{1 , -\}$ \\ [0.5ex] 
 \hline
 \hline
 $Y_{\pi/3}$ &0.747 &0 &0&0.253   \\ 
 \hline
 $X_{\pi/3}$ &0.75 &0.25&0&0  \\
 \hline
 $X_{\pi/3}Y_{\pi/3}$ & 0.56&0.185&0.066&0.188\\
 \hline
 $X_{\pi/3}Y_{2\pi/3}$ & 0.18&0.063&0.184&0.574   \\  
\hline
 $X_{2\pi/3}Y_{\pi/3}$ & 0.19&0.55&0.195&0.063   \\ 
 \hline
 $X_{2\pi/3}Y_{2\pi/3}$ & 0.06&0.19&0.56&0.185 \\
 \hline
  $R=X_{\pi/2}Y_{\pi/2}$&0.25&0.252&0.252&0.245\\
  \hline
  $H$&0&0.503&0.497&0\\[1ex] 
 \hline
 \hline
\end{tabular}
\caption{\textbf{Probability of each type of error.} Here $\{0,+\},\{1,+\},\{0,-\}$ and $\{1,-\}$ represent the two qubit states $\Ket{00},\Ket{10},\Ket{01}$ and $\Ket{11}$ respectively. $+$ is the shorthand for $\Ket{+}=\frac{1}{\sqrt{2}}\big(\Ket{0}+\Ket{1}\big)$ and $-$ is the shorthand for $\Ket{-}=\frac{1}{\sqrt{2}}\big(\Ket{0}-\Ket{1}\big)$. $\Ket{+},\Ket{-}$ are the states of the second ancillary syndrome qubit before the Hadamard operation in the circuit of Fig. \ref{qed_fig2} in the bit-flip and phase-flip cases respectively.}
\label{qed_table4}
\end{table*}
\begin{table*}[h]
\addtolength{\tabcolsep}{20pt}
 \begin{tabular}{|c c c c c|} 
 
 \hline
 \hline
 $\theta$ & $\{0 , +\}$ & $\{1 , +\}$&$\{0 , -\}$&$\{1 , -\}$ \\ [0.5ex] 
 \hline
 \hline
 $-\pi$ &0 &1 &0&0   \\ 
 \hline
 $-14\pi/15$ &0.012 &0.988&0&0  \\
 \hline
 $-13\pi/15$ & 0.045&0.955&0&0\\
 \hline
 $-12\pi/15$ & 0.092&0.908&0&0   \\  
\hline
 $-11\pi/15$ & 0.1644&0.8356&0&0.   \\ 
 \hline
 $-10\pi/15$ & 0.25&0.75&0&0 \\
 \hline
  $-9\pi/15$&0.35&0.65&0&0\\
  \hline
  $-8\pi/15$&0.45&0.55&0&0\\
  \hline
  $-7\pi/15$&0.55&0.45&0&0\\
  \hline
   $-6\pi/15$&0.65&0.35&0&0\\
  \hline
  $-5\pi/15$&0.752&0.248&0&0\\
  \hline
 
  $-4\pi/15$&0.843&0.157&0&0\\
  \hline
  
  $-3\pi/15$&0.905&0.095&0&0\\
  \hline
  $-2\pi/15$&0.952&0.048&0&0\\
  \hline
  $-\pi/15$&0.987&0.013&0&0\\
  \hline
  $0$&1&0&0&0\\
  \hline
  $\pi/15$&0.99&0.0091&0&0\\
  \hline
 $2\pi/15$&0.957&0.043&0&0\\
 \hline
 $3\pi/15$ &0.906 &0.094 &0&0   \\ 
 \hline
 $4\pi/15$ &0.831 &0.17&0&0  \\
 \hline
 $5\pi/15$ & 0.75&0.25&0&0\\
 \hline
 $6\pi/15$ & 0.658&0.342&0&0   \\  
\hline
 $7\pi/15$ & 0.554&0.446&0&0.   \\ 
 \hline
 $8\pi/15$ & 0.436&0.563&0&0 \\
 \hline
  $9\pi/15$&0.34&0.66&0&0\\
  \hline
  $10\pi/15$&0.25&0.75&0&0\\
  \hline
  $11\pi/15$&0.164&0.836&0&0\\
  \hline
  
  $12\pi/15$&0.094&0.91&0&0\\
  \hline
  $13\pi/15$&0.044&0.956&0&0\\
  \hline
 
  $14\pi/15$&0.012&0.988&0&0\\
  \hline
  
  $\pi$&0&1&0&0\\
  [1ex] 
 \hline
 \hline
\end{tabular}
\caption{\textbf{Probability of each type of error for applied error $\varepsilon=X_{\theta}$ with varying $\theta$.}}
\label{qed_table4}
\end{table*}
\begin{table*}[h]
\addtolength{\tabcolsep}{20pt}
 \begin{tabular}{|c c c c c|} 
 
 \hline
 \hline
 $\theta$ & $\{0 , +\}$ & $\{1 , +\}$&$\{0 , -\}$&$\{1 , -\}$ \\ [0.5ex] 
 \hline
 \hline
 $-\pi$ &0 &0 &0&1   \\ 
 \hline
 $-14\pi/15$ &0.011 &0&0&0.99  \\
 \hline
 $-13\pi/15$ & 0.044&0&0&0.956\\
 \hline
 $-12\pi/15$ & 0.098&0&0&0.902   \\  
\hline
 $-11\pi/15$ & 0.166&0&0&0.834   \\ 
 \hline
 $-10\pi/15$ & 0.251&0&0&0.75 \\
 \hline
  $-9\pi/15$&0.35&0&0&651\\
  \hline
  $-8\pi/15$&0.45&0&0&554\\
  \hline
  $-7\pi/15$&0.56&0&0&0.44\\
  \hline
  
  $-6\pi/15$&0.66&0&0&0.34\\
  \hline
  $-5\pi/15$&0.75&0&0&0.25\\
  \hline
 
  $-4\pi/15$&0.84&0&0&0.164\\
  \hline
  
  $-3\pi/15$&0.905&0&0&0.095\\
  \hline
  $-2\pi/15$&0.957&0&0&0.042\\
  \hline
  $-\pi/15$&0.988&0&0&0.012\\
  \hline
  $0$&1&0&0&0\\
  \hline
  $\pi/15$&0.989&0&0&0.011\\
  \hline
 $2\pi/15$&0.957&0&0&0.043\\\hline$3\pi/15$ &0.905 &0 &0&0.095   \\ 
 \hline
 $4\pi/15$ &0.831 &0&0&0.17  \\
 \hline
 $5\pi/15$ & 0.751&0&0&0.25\\
 \hline
 $6\pi/15$ & 0.65&0&0&0.35   \\  
\hline
 $7\pi/15$ & 0.56&0&0&0.44   \\ 
 \hline
 $8\pi/15$ & 0.45&0&0&0.552 \\
 \hline
  $9\pi/15$&0.35&0&0&0.65\\
  \hline
  $10\pi/15$&0.25&0&0&0.75\\
  \hline
  $11\pi/15$&0.168&0&0&0.832\\
  \hline
  
  $12\pi/15$&0.092&0&0&0.908\\
  \hline
  $13\pi/15$&0.039&0&0&0.96\\
  \hline
 
  $14\pi/15$&0.012&0&0&0.99\\
  \hline
  
  $\pi$&0&0&0&1\\
  [1ex] 
 \hline
 \hline
\end{tabular}
\caption{\textbf{Probability of each type of error for applied error $\varepsilon=Y_{\theta}$ with varying $\theta$.}}
\label{qed_table4}
\end{table*}

\begin{table*}[h]
\addtolength{\tabcolsep}{20pt}
 \begin{tabular}{|c c c c c|} 
 
 \hline
 \hline
 $\theta$ & $\{0 , +\}$ & $\{1 , +\}$&$\{0 , -\}$&$\{1 , -\}$ \\ [0.5ex] 
 \hline
 \hline
 $-\pi$ &0 &0 &1&0   \\ 
 \hline
 $-14\pi/15$ &0.011 &0&0.988&0  \\
 \hline
 $-13\pi/15$ & 0.044&0&0.956&0\\
 \hline
 $-12\pi/15$ & 0.096&0&0.904&0   \\  
\hline
 $-11\pi/15$ & 0.163&0&0.837&0   \\ 
 \hline
 $-10\pi/15$ & 0.25&0&0.75&0 \\
 \hline
  $-9\pi/15$&0.35&0&0.65&0\\
  \hline
  $-8\pi/15$&0.45&0&0.55&0\\
  \hline
  $-7\pi/15$&0.55&0&0.45&0\\
  \hline
  
  $-6\pi/15$&0.65&0&0.35&0\\
  \hline
  $-5\pi/15$&0.75&0&0.25&0\\
  \hline
 
  $-4\pi/15$&0.83&0&0.17&0\\
  \hline
  
  $-3\pi/15$&0.91&0&0.09&0\\
  \hline
  $-2\pi/15$&0.96&0&0.04&0\\
  \hline
  $-\pi/15$&0.99&0&0.011&0\\
  \hline
  $0$&1&0&0&0\\
  \hline
  $\pi/15$&0.99&0&0.01&0\\
  \hline
 $2\pi/15$&0.96&0&0.043&0\\\hline $3\pi/15$ &0.903 &0 &0.097&0   \\ 
 \hline
 $4\pi/15$ &0.831 &0&0.17&0  \\
 \hline
 $5\pi/15$ & 0.76&0&0.24&0\\
 \hline
 $6\pi/15$ & 0.65&0&0.34&0   \\  
\hline
 $7\pi/15$ & 0.55&0&0.45&0   \\ 
 \hline
 $8\pi/15$ & 0.44&0&0.56&0 \\
 \hline
  $9\pi/15$&0.35&0&0.65&0\\
  \hline
  $10\pi/15$&0.25&0&0.75&0\\
  \hline
  $11\pi/15$&0.17&0&0.83&0\\
  \hline
  
  $12\pi/15$&0.099&0&0.9&0\\
  \hline
  $13\pi/15$&0.04&0&0.96&0\\
  \hline
 
  $14\pi/15$&0.011&0&0.989&0\\
  \hline
  
  $\pi$&0&0&1&0\\
  [1ex] 
 \hline
 \hline
\end{tabular}
\caption{\textbf{Probability of each type of error for applied error $\varepsilon=Z_{\theta}$ with varying $\theta$.}}
\label{qed_table4}
\end{table*}

\begin{thebibliography}{0}%
\makeatletter
\providecommand \@ifxundefined [1]{%
 \@ifx{#1\undefined}
}%
\providecommand \@ifnum [1]{%
 \ifnum #1\expandafter \@firstoftwo
 \else \expandafter \@secondoftwo
 \fi
}%
\providecommand \@ifx [1]{%
 \ifx #1\expandafter \@firstoftwo
 \else \expandafter \@secondoftwo
 \fi
}%
\providecommand \natexlab [1]{#1}%
\providecommand \enquote  [1]{``#1''}%
\providecommand \bibnamefont  [1]{#1}%
\providecommand \bibfnamefont [1]{#1}%
\providecommand \citenamefont [1]{#1}%
\providecommand \href@noop [0]{\@secondoftwo}%
\providecommand \href [0]{\begingroup \@sanitize@url \@href}%
\providecommand \@href[1]{\@@startlink{#1}\@@href}%
\providecommand \@@href[1]{\endgroup#1\@@endlink}%
\providecommand \@sanitize@url [0]{\catcode `\\12\catcode `\$12\catcode
  `\&12\catcode `\#12\catcode `\^12\catcode `\_12\catcode `\%12\relax}%
\providecommand \@@startlink[1]{}%
\providecommand \@@endlink[0]{}%
\providecommand \url  [0]{\begingroup\@sanitize@url \@url }%
\providecommand \@url [1]{\endgroup\@href {#1}{\urlprefix }}%
\providecommand \urlprefix  [0]{URL }%
\providecommand \Eprint [0]{\href }%
\providecommand \doibase [0]{http://dx.doi.org/}%
\providecommand \selectlanguage [0]{\@gobble}%
\providecommand \bibinfo  [0]{\@secondoftwo}%
\providecommand \bibfield  [0]{\@secondoftwo}%
\providecommand \translation [1]{[#1]}%
\providecommand \BibitemOpen [0]{}%
\providecommand \bibitemStop [0]{}%
\providecommand \bibitemNoStop [0]{.\EOS\space}%
\providecommand \EOS [0]{\spacefactor3000\relax}%
\providecommand \BibitemShut  [1]{\csname bibitem#1\endcsname}%
\let\auto@bib@innerbib\@empty
%</preamble>
\end{thebibliography}%


\clearpage

\begin{thebibliography}{10}
\expandafter\ifx\csname url\endcsname\relax
  \def\url#1{\texttt{#1}}\fi
\expandafter\ifx\csname urlprefix\endcsname\relax\def\urlprefix{URL }\fi
\providecommand{\bibinfo}[2]{#2}
\providecommand{\eprint}[2][]{\url{#2}}
\bibitem{qed_PreskillPhysicstoday1999}
\bibinfo{author}{J.Preskill}   \
\newblock \bibinfo{title}{Battling decoherence: the fault tolerant quantum computer.}
\newblock \emph{\bibinfo{journal}{Phys. Today}} \textbf{\bibinfo{volume}{52}}, \bibinfo{pages}{24--30}
  (\bibinfo{year}{1999}).
\bibitem{qed_ShorIEEE1996}
\bibinfo{author}{Shor, P. W.} 
\newblock \bibinfo{title}{Fault-tolerant quantum computation}.
\newblock \emph{\bibinfo{journal}{{Proc. 37th Conf. Found. of Comp. Sci.}}
}
  (\bibinfo{year}{1996}).
  \bibitem{qed_SteaneNat1999}
\bibinfo{author}{Steane, A.~ M.}
\newblock \bibinfo{title}{Efficient fault-tolerant quantum computing}.
\newblock \emph{\bibinfo{journal}{Nature}} \textbf{\bibinfo{volume}{399}},
  \bibinfo{pages}{124–126} (\bibinfo{year}{1999}).

\bibitem{qed_NiggScience2016}
\bibinfo{author}{Muller, M.} \bibinfo{author}{Rivas, A.} \bibinfo{author}{Martínez, E.~ A.}\bibinfo{author}{Nigg, D.}\bibinfo{author}{Schindler, P.}\bibinfo{author}{Monz, T.}\bibinfo{author}{Blatt, R.}\bibinfo{author}{Martin-Delgado, M.~ A.}
\newblock \bibinfo{title}{Iterative phase optimization of elementary quantum error correcting codes.}
\newblock \emph{\bibinfo{journal}{Phys. Rev. X}} \textbf{\bibinfo{volume}{6}}, \bibinfo{pages}{031030}
  (\bibinfo{year}{2016}).
\bibitem{qed_GhoshQIP2018}
\bibinfo{author}{Ghosh, D.}, \bibinfo{author}{Agarwal, P.},    \bibinfo{author}{Pandey, P.}, \bibinfo{author}{Behera, B.~K.} \& \bibinfo{author}{Panigrahi, P.~K.} \
\newblock \bibinfo{title}{Automated error correction in IBM quantum computer and explicit generalization}.
\newblock \emph{\bibinfo{journal}{Quantum Inf. Process.}} \textbf{\bibinfo{volume}{17}}, \bibinfo{pages}{153}
  (\bibinfo{year}{2018}).

\bibitem{qed_ChiaveriniNature2004}
\bibinfo{author}{Chiaverini, J.}, \bibinfo{author}{Leibfried, D.},
  \bibinfo{author}{Schaetz, T.} \& \bibinfo{author}{Barett, M. D.} 
\newblock \bibinfo{title}{Realization of quantum error correction}.
  \newblock \emph{\bibinfo{journal}{Nature}} 
  \textbf{\bibinfo{volume}{432}},
  \bibinfo{pages}{602--605}
  (\bibinfo{year}{2004}).
  \bibitem{qed_GottesmanarXiv:97050521997}
\bibinfo{author}{Gottesman, D.}
\newblock \bibinfo{title}{Stabilizer Codes and Quantum Error Correction}
\newblock \emph{\bibinfo{journal}{arXiv preperint 	arXiv:quant-ph/9705052 }}
 (\bibinfo{year}{1997}).
\bibitem{qed_RaussendorfPRL2007}
\bibinfo{author}{Raussendorf, R.} \& \bibinfo{author}{Harrington, J.}
\newblock \bibinfo{title}{Fault-tolerant quantum computation with high threshold in two dimensions}.
  \newblock \emph{\bibinfo{journal}{Phys. Rev. Lett.}} 
  \textbf{\bibinfo{volume}{98}},
  \bibinfo{pages}{190504}
  (\bibinfo{year}{2007}).
 \bibitem{qed_ChildressPRL2006}
  \bibinfo{author}{Childress, J.}, \bibinfo{author}{Taylor, J. M.},
  \bibinfo{author}{Sorensen, A. S.} \& \bibinfo{author}{Lukin, M. D.}
\newblock \bibinfo{title}{Fault-tolerant quantum communication based on solid-state photon emitters}.
  \newblock \emph{\bibinfo{journal}{Phys. Rev. Lett.}} 
  \textbf{\bibinfo{volume}{96}},
  \bibinfo{pages}{070504}
  (\bibinfo{year}{2006}).
\bibitem{qed_KnillPRL2001}
\bibinfo{author}{Knill, E.}, \bibinfo{author}{Laflamme, R.},
  \bibinfo{author}{Martinez, R.} \& \bibinfo{author}{Negrevergne, C.}
\newblock \bibinfo{title}{Benchmarking quantum
computers: The five-qubit error correcting code}.
  \newblock \emph{\bibinfo{journal}{Phys. Rev. Lett.}} 
  \textbf{\bibinfo{volume}{86}},
  \bibinfo{pages}{5811–5814}
  (\bibinfo{year}{2001}).
 
\bibitem{qed_Niggscience2014}
\bibinfo{author}{Nigg, D.}\bibinfo{author}{Müller,  M.}\bibinfo{author}{ Martinez,  E.~A.}\bibinfo{author}{ Schindler, P.}\bibinfo{author}{ Hennrich,  M.}\bibinfo{author}{ Monz,  T.}\bibinfo{author}{ Martin-Delgado, M.~A.}\bibinfo{author}{Blatt, R.}
\newblock \bibinfo{title}{Quantum computations on a topologically encoded
qubit.}
\newblock \emph{\bibinfo{journal}{Science}}
  \textbf{\bibinfo{volume}{345}}, \bibinfo{pages}{302--305}
  (\bibinfo{year}{2014}).
\bibitem{qed_LinkeScienceAdvances2017}
\bibinfo{author}{ Linke, N.~M.} \emph{et~al.}
\newblock \bibinfo{title}{Fault-tolerant quantum error detection.}
\newblock \emph{\bibinfo{journal}{Sci. Adv.}}
  \textbf{\bibinfo{volume}{3}},
  \bibinfo{pages}{10}
  (\bibinfo{year}{2017}).

\bibitem{qed_TakitaPRL2017}
\bibinfo{author}{ Takita, M.}, \bibinfo{author}{ Cross, A.~W.},
  \bibinfo{author}{ Córcoles, A. D.}, \ \bibinfo{author}{Chow, J.~M.} \& \bibinfo{author}{Gambetta. J.~M.}
\newblock \bibinfo{title}{Experimental demonstration of fault-tolerant state preparation with
superconducting qubits.}
\newblock \emph{\bibinfo{journal}{Phys. Rev. Lett.}}
  \textbf{\bibinfo{volume}{119}},
  \bibinfo{pages}{180501}
  (\bibinfo{year}{2017}).

\bibitem{qed_FarkasIEEE1995}
\bibinfo{author}{ Farkas, P.}
\newblock \bibinfo{title}{Phase matching condition for quantum search with a
  generalized initial state}.
\newblock \emph{\bibinfo{journal}{IEEE/ACM Trans. Network.}}
  \textbf{\bibinfo{volume}{3}}, \bibinfo{pages}{2}
  (\bibinfo{year}{1995}).
  
\bibitem{qed_FeldmeierIEEE1995}
\bibinfo{author}{Feldmeier, D. ~C.}
\newblock \bibinfo{title}{Fast software implementation of error detection codes}.
\newblock \emph{\bibinfo{journal}{IEEE/ACM Trans. Network.}} \textbf{\bibinfo{volume}{3}}, \bibinfo{pages}{6}
  (\bibinfo{year}{1995}).

\bibitem{qed_McAuleyIEEE1994}
\bibinfo{author}{ McAuley, A. J.}
\newblock \bibinfo{title}{Quantum counting}.
\newblock \emph{\bibinfo{journal}{IEEE/ACM Trans. Network.}}, \textbf{\bibinfo{volume}{2}}, \bibinfo{pages}{1}
  (\bibinfo{year}{1994}).
 
\bibitem{qed_NguyenCISS1998}
\bibinfo{author}{Nguyen, G. D.}
\newblock \bibinfo{title}{A General Class of Error-Detection Codes}.
    \newblock  \emph{\bibinfo{journal}{Proc. 32nd Conf. Inf. Sci. and Sys.}}
   \textbf{\bibinfo{volume}{10}}, 
    \bibinfo{pages}{451--453} (\bibinfo{year}{1998}).
\bibitem{qed_GuptaIJQI2005}
\bibinfo{author}{Gupta, M.}, \bibinfo{author}{Pathak, A.}, \bibinfo{author}{Srikanth, R.} \& \bibinfo{author}{Panigrahi, P.~K.}
\newblock \bibinfo{title}{General circuits for indirecting and distributing measurement in quantum computation}.
\newblock \emph{\bibinfo{journal}{Int. J. Quantum. Inf. }}
  \textbf{\bibinfo{volume}{5}}, \bibinfo{pages}{4}
  (\bibinfo{year}{2005}).
\bibitem{qed_ShorPRA1995}
\bibinfo{author}{Shor, P. W.} \&  \bibinfo{author}{Kenigsberg, D.}
\newblock \bibinfo{title}{Scheme for reducing decoherence in quantum computer memory}.
\newblock \emph{\bibinfo{journal}{{Phys. Rev. A}}
}
  \textbf{\bibinfo{volume}{52}}, \bibinfo{pages}{2493--2496}
  (\bibinfo{year}{1995}).


\bibitem{qed_SteanePRL1996}
\bibinfo{author}{Steane, A.~ M.}
\newblock \bibinfo{title}{Error Correcting Codes in Quantum Theory}.
\newblock \emph{\bibinfo{journal}{Phys. Rev. Lett.}} \textbf{\bibinfo{volume}{77}},
  \bibinfo{pages}{793--797} (\bibinfo{year}{1996}).

\bibitem{qed_SteanePRSL1996}
\bibinfo{author}{Steane, A. } 
\newblock \bibinfo{title}{ Multiple particle interference and quantum error correction}.
\newblock \emph{\bibinfo{journal}{Proc. Roy. Soc.
Lond.}}
  \textbf{\bibinfo{volume}{452}}, \bibinfo{pages}{2551}
  (\bibinfo{year}{1996}).

\bibitem{qed_SteaneIEEE1999}
\bibinfo{author}{Steane, A.}
\newblock \bibinfo{title}{Enlargement of Calderbank Shor Steane quantum codes}\newblock \emph{\bibinfo{journal}. {IEEE Trans. on Inf. Theory}}  \textbf{\bibinfo{volume}{45}}, \bibinfo{pages}{2492–2495} (\bibinfo{year}{1999}).
\bibitem{qed_CorcolesNatComm2015}
    \bibinfo{author}{Córcoles, A. D.} \bibinfo{author}{Magesan, E.}\bibinfo{author}{Srinivasan, S.~J.} \bibinfo{author}{Cross, A.~W.} \bibinfo{author}{Steffen, M.} \bibinfo{author}{Gambetta, J.~M.}{Chow, J.~M.}
\newblock \bibinfo{title}{Demonstration of a quantum error detection code using a square lattice of four superconducting qubits.}
\newblock \emph{\bibinfo{journal}{Nat. Commun.}}
  \textbf{\bibinfo{volume}{6}}, \bibinfo{pages}{6979} 
  (\bibinfo{year}{2015}).
  \bibitem{qed_BravyiarXiv9811052}
    \bibinfo{author}{Bravyi, S.} \& \bibinfo{author}{Kitaev A.}
\newblock \bibinfo{title}{Quantum codes on a lattice with boundary.}
\newblock \emph{\bibinfo{journal}{arxiv Preprint arXiv quant-ph/9811052}}
   (\bibinfo{year}{1998}).
   \bibitem{qed_KitaevAP1997}
 \bibinfo{author}{Kitaev A.}
\newblock \bibinfo{title}{Fault-tolerant quantum computation by anyons.}
\newblock \emph{\bibinfo{journal}{Ann. Phys.}}
   \textbf{\bibinfo{volume}{303}}, 
   \bibinfo{pages}{2--30}
   (\bibinfo{year}{1997}).
\bibitem{qed_FowlerPRA2012}
\bibinfo{author}{Fowler, A. G.}, \bibinfo{author}{Mariantoni, M.},
  \bibinfo{author}{Martinis, J. M.} \& \bibinfo{author}{Cleland, A. N.} 
\newblock \bibinfo{title}{Surface codes: towards practical large-scale quantum computation.}
  \newblock \emph{\bibinfo{journal}{Phys. Rev. A }} 
  \textbf{\bibinfo{volume}{86}},
  \bibinfo{pages}{032324}
  (\bibinfo{year}{2012}).  
\bibitem{qed_ChowNatcommun2014}
\bibinfo{author}{Chow, J. M.}\bibinfo{author}{Gambetta, J.M.}\bibinfo{author}{Magesan, E.}\bibinfo{author}{Abraham, D.W.}\bibinfo{author}{Cross, A.W.}\bibinfo{author}{Johnson, B.R.}\bibinfo{author}{Masluk, N.A.} \bibinfo{author}{Ryan, C.A.}\bibinfo{author}{Smolin, J.A.}\bibinfo{author}{Srinivasan, S.J.}\bibinfo{author}{M Steffen, M.}
\newblock \bibinfo{title}{Implementing a strand of a scalable fault-tolerant quantum computing fabric.}
\newblock \emph{\bibinfo{journal}{Nat. Commun.}}
  \textbf{\bibinfo{volume}{5}}, 
  \bibinfo{pages}{500}
  (\bibinfo{year}{2014}).
\bibitem{qed_PaikPRL2011}
\bibinfo{author}{Paik, H.}\bibinfo{author}{Schuster, D.~I.}\bibinfo{author}{ Bishop, L.~S.}\bibinfo{author}{Kirchmair, G.}\bibinfo{author}{Catelani, G.}\bibinfo{author}{Sears, A.~P.}\bibinfo{author}{Johnson, B.~R.}\bibinfo{author}{Reagor, M.J.}\bibinfo{author}{Frunzio, L.}\bibinfo{author}{Glazman, L.}\bibinfo{author} {Girvin, S.~M.}\bibinfo{author}{Devoret, M.~H.}\bibinfo{author}{Schoelkopf, R.~J.}
\newblock \bibinfo{title}{ Observation of high coherence in josephson junction qubits measured in a three-dimensional circuit qed architecture.}
\newblock \emph{\bibinfo{journal}{Phys. Rev. Lett.}}
  \textbf{\bibinfo{volume}{107}}, 
  \bibinfo{pages}{240501}
  (\bibinfo{year}{2011}).
\bibitem{qed_ChuangAPL2013}
\bibinfo{author}{Chang, J.}\bibinfo{author}{Vissers, M.R.}\bibinfo{author}{Corcoles, A.D.}\bibinfo{author}{Sandberg, M.}\bibinfo{author}{Gao, J.}\bibinfo{author}{Abraham, D.W.}\bibinfo{author}{Chow, J.M.}\bibinfo{author}{Jay M. Gambetta, J}\bibinfo{author}{Rothwell, M.B.}\bibinfo{author}{Keefe, G.A.}\bibinfo{author}{Steffen, M.}\bibinfo{author}{Pappas, D.P.}
\newblock \bibinfo{title}{Improved superconducting qubit coherence using titanium nitride.}
\newblock \emph{\bibinfo{journal}{Appl. Phys. Lett.}}
  \textbf{\bibinfo{volume}{103}}, 
  \bibinfo{pages}{012602}
  (\bibinfo{year}{2013}).
\bibitem{qed_BarendaPRL2013}
\bibinfo{author}{Barends, R.}\bibinfo{author}{Kelly, J.}\bibinfo{author}{Megrant, A.} \bibinfo{author}{Sank, D.}\bibinfo{author}{Jeffrey, E.}\bibinfo{author}{Chen, Y.}\bibinfo{author}{Yin, Y.}\bibinfo{author}{Chiaro, B.}\bibinfo{author}{Mutus, J.}\bibinfo{author}{Neill, C.}\bibinfo{author}{O’Malley, P.}\bibinfo{author}{Roushan, P.}\bibinfo{author}{Wenner, J.}\bibinfo{author}{White, T.~C.}\bibinfo{author}{White, T.~C.}\bibinfo{author}{Martinis, J.~M.}
\newblock \bibinfo{title}{ Coherent josephson qubit suitable for scalable quantum integrated circuits.}
\newblock \emph{\bibinfo{journal}{Phys. Rev. Lett.}}
  \textbf{\bibinfo{volume}{111}}, 
  \bibinfo{pages}{080502}
  (\bibinfo{year}{2013}).    
\bibitem{qed_BarendaNat2014}
\bibinfo{author}{Barends, R.}\bibinfo{author}{Kelly, J.}\bibinfo{author}{ Megrant, A.}\bibinfo{author}{Veitia, A.}\bibinfo{author}{Sank, D.}\bibinfo{author}{Jeffrey, E.}\bibinfo{author}{ White, T.C.}\bibinfo{author}{ Mutus,  J.}\bibinfo{author}{Fowler, A.G.}\bibinfo{author}{ Campbell, B.}\bibinfo{author}{Chen, Y.}\bibinfo{author}{ Chen, Z.}\bibinfo{author}{ Chiaro, B.}\bibinfo{author}{ Dunsworth, A.}\bibinfo{author}{ Neill, C.}\bibinfo{author}{O'Malley, P.}\bibinfo{author}{Roushan, P.}\bibinfo{author}{ Vainsencher, A.}\bibinfo{author}{Wenner, J.}\bibinfo{author}{Korotkov, A.N.}\bibinfo{author}{Cleland, A.N.}\bibinfo{author}\bibinfo{author}{Martinis, J.M.}
\newblock \bibinfo{title}{Superconducting quantum circuits at the surface code threshold for fault tolerance.}
\newblock \emph{\bibinfo{journal}{Nature}}
  \textbf{\bibinfo{volume}{508}}, 
  \bibinfo{pages}{500--503}
  (\bibinfo{year}{2014}). 


\bibitem{qed_AggarwalarXiv:1804.08655v12018}
\bibinfo{author}{Aggarwal, D.},  \bibinfo{author}{Raj, S.}, \bibinfo{author}{Behera, B.~K.} \& \bibinfo{author}{Panigrahi, P.~K.} 
\newblock \bibinfo{title}{Application of quantum scrambling in Rydberg atom on IBM quantum computer}.
\newblock \emph{\bibinfo{journal}{arXiv preprint
arXiv:1804.08655v1}} (\bibinfo{year}{2018}).

\bibitem{qed_SrinivasanarXiv:1805.109282018}
\bibinfo{author}{ Srinivasan, K.},  \bibinfo{author}{ Satyajit, S.}, \bibinfo{author}{Behera, B.~K.}  \& \bibinfo{author}{Panigrahi, P.~K.} 
\newblock \bibinfo{title}{Efficient quantum algorithm for solving travelling salesman problem: An IBM quantum experience}.
\newblock \emph{\bibinfo{journal}{arXiv preprint arXiv:1805.10928}} (\bibinfo{year}{2018}).

\bibitem{qed_BeheraQIP2017}
\bibinfo{author}{Behera, B.~K.}, \bibinfo{author}{Banerjee, A.} \& \bibinfo{author}{Panigrahi, P.~K.}
\newblock \bibinfo{title}{Experimental realization of quantum cheque using a five-qubit quantum computer}.
\newblock \emph{\bibinfo{journal}{Quantum Inf. Process.}}
  \textbf{\bibinfo{volume}{16(12)}}, \bibinfo{pages}{312}
  (\bibinfo{year}{2017}).
 \bibitem{qed_DasharXiv1710.051962017}
\bibinfo{author}{Dash, A.}, \bibinfo{author}{Rout, S.}, \bibinfo{author}{Behera, B.~K.} \& \bibinfo{author}{Panigrahi, P.~K.}
\newblock \bibinfo{title}{A Verification Algorithm and Its Application to Quantum Locker in IBM Quantum Computer}.
\newblock \emph{\bibinfo{journal}{arXiv preprint arXiv:1710.05196}}
   (\bibinfo{year}{2017}).
   \bibitem{qed_VishnuarXiv:1709.05697}
\bibinfo{author}{Vishnu, P. K.}, \bibinfo{author}{Joy, D.}, \bibinfo{author}{Behera, B.~K.} \& \bibinfo{author}{Panigrahi, P.~K.}
\newblock \bibinfo{title}{Experimental Demonstration of Non-local Controlled-Unitary Quantum Gates Using a Five-qubit Quantum Computer}.
\newblock \emph{\bibinfo{journal}{arXiv preprint arXiv:1709.05697}}
   (\bibinfo{year}{2017}). 
\bibitem{qed_SatyajitarXiv:1712.05485}
\bibinfo{author}{Satyajit, S.}, \bibinfo{author}{Srinivasan, K.}, \bibinfo{author}{Behera, B.~K.} \& \bibinfo{author}{Panigrahi, P.~K.}
\newblock \bibinfo{title}{Discrimination of Highly Entangled Z-states in IBM Quantum Computer}.
\newblock \emph{\bibinfo{journal}{arXiv preprint arXiv:1712.05485}}
   (\bibinfo{year}{2017}). 
\bibitem{qed_RoyarXiv:1710.10717}
\bibinfo{author}{Roy, S.}, \bibinfo{author}{Behera, B.~K.} \& \bibinfo{author}{Panigrahi, P.~K.}
\newblock \bibinfo{title}{Demonstration of Entropic Noncontextual Inequality Using IBM Quantum Computer}.
\newblock \emph{\bibinfo{journal}{arXiv preprint arXiv:1710.10717}}
   (\bibinfo{year}{2017}). 

\bibitem{qed_GangopadhyayQIP2017}
\bibinfo{author}{ Gangopadhyay, S.}, \bibinfo{author}{Manabputra.} \bibinfo{author}{Behera, B.~K.} \& \bibinfo{author}{Panigrahi, P.~K.}
\newblock \bibinfo{title}{Generalization and demonstration of an entanglement-based Deutsch–Jozsa-like algorithm using a 5-qubit quantum computer}.
\newblock \emph{\bibinfo{journal}{Quantum Inf. Process.}} \textbf{\bibinfo{volume}{17}}, \bibinfo{pages}{160}
 (\bibinfo{year}{2017}). 

\bibitem{qed_HegadearXiv:1712.073262017}
\bibinfo{author}{Hegade, N. N.}, \bibinfo{author}{Behera, B.~K.}
  \& \bibinfo{author}{Panigrahi, P.~K.}
  \newblock \bibinfo{title}{Experimental Demonstration of Quantum Tunneling in IBM Quantum Computer)}.
\newblock \emph{\bibinfo{journal}{arXiv preprint arXiv:1712.07326}}
(\bibinfo{year}{2017}).
\bibitem{qed_BeheraarXiv:1712.008542017}
\bibinfo{author}{Behera, B.~K.}, \bibinfo{author}{Seth, S.},   \bibinfo{author}{Das, A.} \& \bibinfo{author}{Panigrahi, P.~K.}  
\newblock \bibinfo{title}{Experimental Demonstration of Quantum Repeater in IBM Quantum Computer}.
\newblock \emph{\bibinfo{journal}{arXiv preprint arXiv:1712.00854}}
(\bibinfo{year}{2017}).
\bibitem{qed_KalraarXiv:1707.09462}
\bibinfo{author}{Kalra, A. R.}, \bibinfo{author}{Prakash, S.}, \bibinfo{author}{Behera, B.~K.} \& \bibinfo{author}{Panigrahi, P.~K.}
\newblock \bibinfo{title}{Experimental Demonstration of the No Hiding Theorem Using a 5 Qubit Quantum Computer}.
\newblock \emph{\bibinfo{journal}{arXiv preprint arXiv:1707.09462}}
   (\bibinfo{year}{2017}).
   \bibitem{qed_JhaarXiv:1806.10221}
\bibinfo{author}{Jha, R.} \bibinfo{author}{Das,D.}\bibinfo{author}{Dash, D.}\bibinfo{author}{Jayaraman, S.}\bibinfo{author}{Behera, B.K.}\bibinfo{author}{Panigrahi, P.K.}
\newblock \bibinfo{title}{A Novel Quantum N-Queens Solver Algorithm and its Simulation and Application to Satellite Communication Using IBM Quantum Experience}.
\newblock \emph{\bibinfo{journal}{arXiv preprint arXiv:1806.10221}}
  (\bibinfo{year}{2018}).
 
\bibitem{qed_DasharXiv:1805.10478}
\bibinfo{author}{Dash, A.}, \bibinfo{author}{Sarmah, D.} \bibinfo{author}{Behera, B.~K.}
  \& \bibinfo{author}{Panigrahi, P.~K.}
  \newblock \bibinfo{title}{Exact search algorithm to factorize large biprimes and a triprime on IBM quantum computer}.
\newblock \emph{\bibinfo{journal}{arXiv preprint arXiv:1805.10478}}
(\bibinfo{year}{2018}).
\bibitem{qed_BeherarXiv:1803.06530}
\bibinfo{author}{Behera, B.~K.} \bibinfo{author}{Reza, T.} \bibinfo{author}{Gupta, A.}
  \& \bibinfo{author}{Panigrahi, P.~K.}
  \newblock \bibinfo{title}{Designing Quantum Router in IBM Quantum Computer}.
\newblock \emph{\bibinfo{journal}{arXiv preprint arXiv:1803.06530}}
(\bibinfo{year}{2018}).
\bibitem{qed_SrinivasanarXiv:1801.00778}
 \bibinfo{author}{Srinivasan, K.}, \bibinfo{author}{Behera, B.~K.} 
  \& \bibinfo{author}{Panigrahi, P.~K.}
  \newblock \bibinfo{title}{Solving Linear Systems of Equations by Gaussian Elimination Method Using Grover's Search Algorithm: An IBM Quantum Experience}.
\newblock \emph{\bibinfo{journal}{arXiv preprint arXiv:1801.00778}}
(\bibinfo{year}{2018}).
\bibitem{qed_GurnaniarXiv:1712.10231}
 \bibinfo{author}{Gurnani, K.} \bibinfo{author}{Behera, B.~K.} 
  \& \bibinfo{author}{Panigrahi, P.~K.}
  \newblock \bibinfo{title}{Demonstration of Optimal Fixed-Point Quantum Search Algorithm in IBM Quantum Computer}.
\newblock \emph{\bibinfo{journal}{arXiv preprint arXiv:1712.10231}}
(\bibinfo{year}{2017}).

\bibitem{qed_KapilarXiv:1807.00521}
\bibinfo{author}{Kapil, M.}, \bibinfo{author}{Behera, B.~K.} 
  \& \bibinfo{author}{Panigrahi, P.~K.}
  \newblock \bibinfo{title}{Quantum Simulation of Klein Gordon Equation and Observation of Klein Paradox in IBM Quantum Computer}.
\newblock \emph{\bibinfo{journal}{arXiv preprint arXiv:1807.00521}}
(\bibinfo{year}{2018}).
\bibitem{qed_MohantaarXiv:1807.00323}
\bibinfo{author}{Mohanta, Y. M.} \emph{et~al.}  \
\newblock \bibinfo{title}{Spin-Boson Model to Demonstrate Quantum Tunneling in Biomolecules using IBM Quantum Computer}.
\newblock \emph{\bibinfo{journal}{arXiv preprint arXiv:1807.00323}}
  (\bibinfo{year}{2018}).
\bibitem{qed_BeherarXiv:1806.10229}
\bibinfo{author}{Manabputra}, \bibinfo{author}{Behera, B.~K.} \& \bibinfo{author}{Panigrahi, P.~K.}
  \newblock \bibinfo{title}{A Simulational Model for Witnessing Quantum Effects of Gravity Using IBM Quantum Computer}.
\newblock \emph{\bibinfo{journal}{arXiv preprint arXiv:1806.10229}}
(\bibinfo{year}{2018}).


\bibitem{qed_HarperarXiv:1806.023592018}
\bibinfo{author}{Harper, R.} \& \bibinfo{author}{Flammia, S.}
\newblock \bibinfo{title}{Fault tolerance in the IBM Q Experience}.
\newblock \emph{\bibinfo{journal}{arXiv preprint arXiv:1806.02359}}
(\bibinfo{year}{2018}). 

\bibitem{qed_KlcoarXiv:1803.033262018}
\bibinfo{author}{Klco, N.}\bibinfo{author}{Dumitrescu, E.F.}\bibinfo{author}{McCaskey, A.J.}\bibinfo{author}{Morris, T.D.}\bibinfo{author}{Pooser, R.C.}\bibinfo{author}{Sanz, M.}\bibinfo{author}{Solano, E.}\bibinfo{author}{Lougovski, P.}\bibinfo{author}{Savage, M.J.}
\newblock \bibinfo{title}{Quantum-Classical Computations of Schwinger Model Dynamics using Quantum Computers}.
\newblock \emph{\bibinfo{journal}{arXiv preprint arXiv:1803.03326}} (\bibinfo{year}{2018}).

\bibitem{qed_VuillotarXiv:1705.089572017}
\bibinfo{author}{Vuillot, C.}
\newblock \bibinfo{title}{Is error detection helpful on IBM 5Q chips?}
\newblock \emph{\bibinfo{journal}{arXiv preprint arXiv:1705.08957}} (\bibinfo{year}{2017}).
\bibitem{qed_BennettPRL1993}
\bibinfo{author}{Bennett, C.H.}\bibinfo{author}{Brassard, G.}\bibinfo{author}{Crépeau, C.}\bibinfo{author}{Jozsa, R.}\bibinfo{author}{Peres, A.}\bibinfo{author}{Wootters, W.K.}
\newblock \bibinfo{title}{Teleporting an Unknown Quantum State via Dual Classical and EPR Channels}.
\newblock \emph{\bibinfo{journal}{Phys. Rev. Lett.}} 
\textbf{\bibinfo{volume}{70}},
\bibinfo{pages}{1895--1899}
(\bibinfo{year}{1993}).
\bibitem{qed_BennettIEEE1984}
\bibinfo{author}{Bennett, C.H.} \emph{et~al.}
\newblock \bibinfo{title}{Quantum cryptography: Public key distribution and coin tossing}.
\newblock \emph{\bibinfo{journal}{Proc. IEEE Int. Conf. Comp. Sys. Sig. Process.}} 
\textbf{\bibinfo{volume}{175}},
\bibinfo{pages}{8}
(\bibinfo{year}{1984}).
\bibitem{qed_BennettJC1992}
\bibinfo{author}{Bennett, C.H.} \emph{et~al.}
\newblock \bibinfo{title}{Experimental Quantum Cryptography}.
\newblock \emph{\bibinfo{journal}{Jour. Crypt.}} 
\textbf{\bibinfo{volume}{5}},
\bibinfo{pages}{3--28}
(\bibinfo{year}{1992}).
\bibitem{qed_HilleryarXiv98060631998}
\bibinfo{author}{Hillery, M.}, \bibinfo{author}{Buzek, V.} \& \bibinfo{author}{Berthiaume, A.} 
\newblock \bibinfo{title}{Quantum secret sharing}.
\newblock \emph{\bibinfo{journal}{arXiv Preprint 	arXiv:quant-ph/9806063}} 
(\bibinfo{year}{1998}).
\bibitem{qed_BennettPRL1992}
\bibinfo{author}{Bennett, C.H.} \& \bibinfo{author}{Wiesner, S.}  
\newblock \bibinfo{title}{Communication via one-and two-particle operators on Einstein-Podolsky-Rosen states}.
\newblock \emph{\bibinfo{journal}{Phys. Rev. Lett.}} 
\textbf{\bibinfo{volume}{69}},
\bibinfo{pages}{2881}
(\bibinfo{year}{1992}).
\bibitem{pani_ent}
\bibinfo{author}{Bhaskara, V.S.}, \& \bibinfo{author}{Panigrahi, P.K.}  
\newblock \bibinfo{title}{Generalized concurrence measure for faithful quantification of multiparticle pure state entanglement using Lagrange's identity and wedge product}.
\newblock \emph{\bibinfo{journal}{Quant. Inf. Process}} 
\textbf{\bibinfo{volume}{16}},
\bibinfo{pages}{118}
(\bibinfo{year}{2017}).
\bibitem{qed_KochPRA2007}
\bibinfo{author}{Koch, J.}\bibinfo{author}{Yu, T.M.}\bibinfo{author}{Gambetta, J.}\bibinfo{author}{Houck, A.A.}\bibinfo{author}{Schuster, D.I.}\bibinfo{author}{Majer, J.}\bibinfo{author}{Blais, A.}\bibinfo{author}{Devoret, M.H.}\bibinfo{author}{Girvin, S.M.}\bibinfo{author}{Schoelkopf, R.J.}
\newblock \bibinfo{title}{Charge-insensitive qubit design derived from the Cooper pair box}.
\newblock \emph{\bibinfo{journal}{Phys. Rev. A}} \textbf{\bibinfo{volume}{76}},
  (\bibinfo{year}{2007}).

\bibitem{qed_IBM}
\bibinfo{author}{IBM}
\newblock \bibinfo{title}{Quantum Experience, URL:https://www.research.ibm.com/ibm-q/}.


\bibitem{qed_QISKit}
\bibinfo{author}{IBM} 
\newblock \bibinfo{title}{Quantum Information Software Kit (QISKit), URL:https://www.qiskit.org}.

\bibitem{qed_WangarXiv:1801.037822018}
\bibinfo{author}{Wang, Y.}, \bibinfo{author}{Li, Y.}, \bibinfo{author}{Yin, Z.} \& \bibinfo{author}{Zeng, B.}
\newblock \bibinfo{title}{16-qubit IBM universal quantum computer can be fully entangled}.
\newblock \emph{\bibinfo{journal}{arXiv preprint arXiv:1801.03782}} (\bibinfo{year}{2018}).
\bibitem{graphstateent}
Hein, M., Eisert, J. \& Briegel, H. J. Multiparty entanglement in graph states. \textit{Physical Review
A} \textbf{69}, 062311 (2004).
\end{thebibliography}
\end{document}